\documentclass[twocolumn, superscriptaddress, amsmath, amssymb, aps]{revtex4-2}

\usepackage{graphicx}% Include figure files
\usepackage{dcolumn}% Align table columns on decimal point
\usepackage{bm}% bold math
\usepackage{xcolor}
\usepackage{mathtools}
\usepackage{physics}
\usepackage[normalem]{ulem}
\usepackage{soul}
\usepackage{braket}
\usepackage{appendix}
\usepackage{xcolor}   
\usepackage{xr}
\usepackage{array} % for defining new column types
\usepackage{hyperref} % the last package to be imported
\hypersetup{colorlinks=true, citecolor=blue}

\newcommand{\rev}[1]{{\color{black} #1}}  %colors revisions as red, comment out for clean copy
\newcolumntype{P}[1]{>{\centering\arraybackslash}p{#1}}

\begin{document}

\title{Superuniversal Statistics with Topological Origins for non-Hermitian Scattering Singularities}

\author{Nadav Shaibe}
\email{Corresponding author: nshaibe@umd.edu}
 \affiliation{Maryland Quantum Materials Center, Department of Physics, University of Maryland, College Park, Maryland 20742-4111, USA}

\author{Jared M. Erb}
 \affiliation{Maryland Quantum Materials Center, Department of Physics, University of Maryland, College Park, Maryland 20742-4111, USA}
 
\author{Steven M. Anlage}
 \affiliation{Maryland Quantum Materials Center, Department of Physics, University of Maryland, College Park, Maryland 20742-4111, USA}

\date{\today}

\begin{abstract}
Vortex singularities \rev{in} speckle patterns \rev{formed from} random superpositions of waves are an inevitable consequence of destructive interference and are consequently generic and ubiquitous. Singularities are topologically stable, meaning they persist under small perturbations and can only be removed via pairwise annihilation. \rev{They have applications including sensing, imaging and energy transfer in multiple fields such as optics, acoustics, and elastic or fluid waves}. We generalize the concept of singularity speckle patterns to arbitrary two dimensional parameter spaces and any complex scalar function that describes wave phenomena involving complicated scattering. In scattering systems specifically, we are often concerned with singularities associated with complex zeros of \rev{various functions of} the scattering matrix $S$. Some examples are Coherent Perfect Absorption (CPA), Reflectionless Scattering Modes (RSMs), Transmissionless Scattering Modes (TSMs), and $S$-matrix Exceptional Points (EPs). Experimentally, we find that all scattering singularities share a universal statistical property: \rev{any quantity that diverges as a simple pole at a singularity, such as $\text{det}S$ at CPA, has a probability distribution function with a $-3$ power law tail.} The heavy tail of the distribution provides an estimate for the likelihood of finding a given singularity in a generic scattering system.  We use these universal statistical results to determine that homogeneous system loss is the most important parameter determining singularity density in a given parameter space \rev{of an absorptive scattering system}. Finally, we discuss events where distinct singularities coincide in parameter space, which result in higher order singularities that have new applications \rev{beyond the capabilities of isolated singularities}. These higher order singularities are not topologically protected, and we do not find universal statistical properties for them. We support our empirical results from microwave experiments with Random Matrix Theory simulations and conclude that \rev{the statistical} results presented \rev{hold} for all generic \rev{non-Hermitian} scattering systems \rev{in which singularities can occur}.\\
\end{abstract}

\maketitle

\section{Introduction}\label{Sec_Intro}

%\textit{Introduction}.--
Topological defects in superpositions of complex plane waves where the amplitude vanishes were first \rev{formally introduced} by Nye and Berry in 1974 \cite{Nye1974}. They showed these phase dislocations, later called vortices or singularities, are due to interference effects, not dispersion. Vortices are present in any complex scalar field, and come with integer winding numbers which, with parametric evolution of the system, allows for pair-wise creation and annihilation of vortices with opposite winding numbers \cite{Neu1990}. Due to the ubiquity of singularities, they appear and have been studied in both classical and quantum systems \cite{KT73,Bliokh2017}, from speckle patterns in disordered optics \cite{Freund1993,Zhang2007,Cao2022} to nodes of wavefunctions \cite{Hirschfelder1974,Heckenberg1992}, superconductors \cite{Abrikosov57,Beasley79,Song2009,wang2025} and quantum weak measurements \cite{Solli2004}.

In a smooth complex scalar field $\mathcal{S}$ which occupies a $d$-parameter space, there are an infinite number of $d-1$ dimensional structures that are topologically stable, meaning their existence is robust to small perturbations of the field. An example of such a structure is the contour of $\text{Re}[\mathcal{S}]=0$  ($\text{Im}[\mathcal{S}]=0$), although any scalar value that is within the bounds of $\text{Re}[\mathcal{S}]$ ($\text{Im}[\mathcal{S}]$) will have the same stability. The intersection of two distinct $d-1$ dimensional structures of the same field results in topologically stable $d-2$ dimensional structures.  These structures are the vortex singularities that have drawn so much interest. The simplest example is $\mathcal{S}=0+i0$ as the intersection of $\text{Re}[\mathcal{S}]=0$ and $\text{Im}[\mathcal{S}]=0$, but the intersection of $\text{Re}[\mathcal{S}]=a$ and $\text{Im}[\mathcal{S}]=b$ is the complex zero of the field $\mathcal{S}-(a+ib)$ \cite{Hajnal1987,Berry2000,DeAngelis2016}. In this paper, we will use the term ``fundamental singularity'' to refer to a vortex of a complex scalar field, whatever that field may be.

However, while two distinct $d-1$ structures of a single field can intersect to create a $d-2$ structure, the intersection of two $d-2$ structures of a single field results in vortex-antivortex annihilation, so there are no $d-3$ or beyond dimensional structures. This is equivalent to stating that a complex scalar field has two degrees of freedom, $d-1$ structures are constraints on one of those degrees of freedom \rev{(such as constraining only $\text{Re}[\mathcal{S}]$)}, while $d-2$ structures are constraints on both \rev{(i.e. both $\text{Re}[\mathcal{S}]$ and $\text{Im}[\mathcal{S}]$)}. It is impossible to constrain more degrees of freedom than $\mathcal{S}$ has. Hence without loss of generality, we can use the language of the $d=2$ case for simplicity, that being topologically protected points ($d-2$) and curves ($d-1$) \cite{Dennis2025}. In the \rev{rest of this paper}, any mention of points and curves is implicitly referring to ``points in two dimensional parameter space'' and ``curves in two dimensional parameter space''.

A network of such points in two dimensions result in what are known as ``speckle patterns'' \cite{Goodman1975,Goodman1976,Zhang2010}. We reiterate the pervasiveness of singularities by referencing the use and study of speckle patterns for metrology \cite{Fujii1976,Stetson1987,Wang2006}, velocimetry \cite{Barker1977,Andres1999}, non-destructive testing \cite{YANG1995,BRUNO2000}, imaging techniques such as Digital Image Correlation \cite{Dong2017,Crammond2013} and Speckle Contrast Imaging \cite{Briers2013,Heeman2019}, and more. There is also a rich literature about the statistical properties of speckle patterns \cite{Dainty1970,Wagner1983,Okamoto1995,Zhang2007}. In this paper, the concept of speckle patterns is generalized to any complex scalar field in arbitrary parameter space. Singularities as curves in three dimensional parameter space can be seen in Fig. 2 of Ref.~\cite{Zhang2010}, and Appendix \ref{App_3ParaSpace}.

In this paper we specifically consider the singularities of the scattering $S$-matrix of a non-Hermitian scattering cavity that is open to the outside world through $M$ asymptotic scattering channels. The scattering matrix relates \rev{the vectors of} ingoing $\psi_{in}$ and outgoing $\psi_{out}$ monochromatic waves by $\ket{\psi_{out}} = S \ket{\psi_{in}}$, and can be related to the effective Hamiltonian of the closed cavity through the Heidelberg approach \cite{VERBAARSCHOT1998,SOKOLOV1989,Fyodorov1997bTRI,FSav11,Kuhl2013,Schomerus2015}. We focus on the $S$-matrix because it is experimentally accessible in many scattering platforms (including acoustics, particle physics, quantum transport, non-destructive testing, electromagnetic waves, etc.) and contains a depth of information about the closed system and its scattering processes. Related to the $S$-matrix are the impedance $Z$, admittance $Y$, and Wigner reaction $K$ matrices \cite{Hul2005} that have similar properties and show the same singularity statistics.

One method through which information about a system can be extracted from the $S$-matrix is analysis of complex time delays. Complex time delays are energy (or equivalently frequency) derivatives of functions of the $S$-matrix (such as $\text{det}S$ or $S_{xx}$) and can be directly related to time and center-frequency shifts of time domain pulses propagating through the system \cite{Asano2016,giovannelli2025}. Time delays have previously been used to determine the locations of poles and zeros of $S$-matrix in the complex frequency plane \cite{Fyodorov1997bTRI,Fyodorov2017,Fyodorov2019,Osman2020,Lei2021WS,Lei2022} which are the related to the eigenvalues of the system's effective Hamiltonian \cite{Osman2020,Lei2021WS,Lei2022,Ma2023}. Another established application of time delay is identifying optimal wave configurations with high sensitivity, for use in both cavity- and wavefront-shaping \cite{Rotter2011,Gerardin2016,Horodynski2020,Hougne2021}. 

Recently it was found that the probability density functions (PDF) of every complex time delay quantity 
\cite{Fyodorov2019,Lei2022,Patel2023,Mao2023,Saalman2024} in non-Hermitian systems has the same universal $-3$ power law tail \cite{Lei2021Stats,shaibe2025}. Since complex time delays diverge at singularities of their associated scattering parameter \cite{Hougne2021,Huang2022}, the tail of the PDFs give a measure of the abundance of these phenomena. In this paper we will go further to show that the PDF of any quantity that \rev{has a simple pole divergence} solely at a scattering singularity has the same $-3$ power law tail independent of the function, singularity, or system parameters.

\subsection{Fundamental Singularities} \label{Sub_FunSing}
One singularity that has drawn a lot of interest is known as anti-lasing or alternatively as Coherent Perfect Absorption (CPA). This is a phenomena characterized by all the energy injected into a system being completely absorbed, with no reflection or transmission \cite{Chong2010,Fyodorov2017,Lei2020,Imani2020,Frazier2020,Hougne2021,Erb2024,Faul2025}. CPA has been found in optics \cite{Gupta2012,Pichler2019,Baranov2017}, acoustics \cite{Song2014,Meng2017,IglesiasMartinez2025}, heat transfer \cite{Li2022}, and quantum single photon systems \cite{Vetlugin2021,Vetlugin2022_Resolution,Vetlugin2022_AHOM}. In microwave scattering, the signature of CPA is an eigenvalue of the $S$-matrix \rev{achieving complex zero}, so $\text{det}S$ is the complex scalar function who's zero is the enabling condition for CPA \cite{Guo2023}. Alongside enabling CPA,  $\text{det}S=0+i0$ also turns a two channel system into a tunable robust splitter \cite{erb2025robustwavesplittersbased}.

\begin{figure*}[bth]
\centerline{
\includegraphics[width=19.2cm]{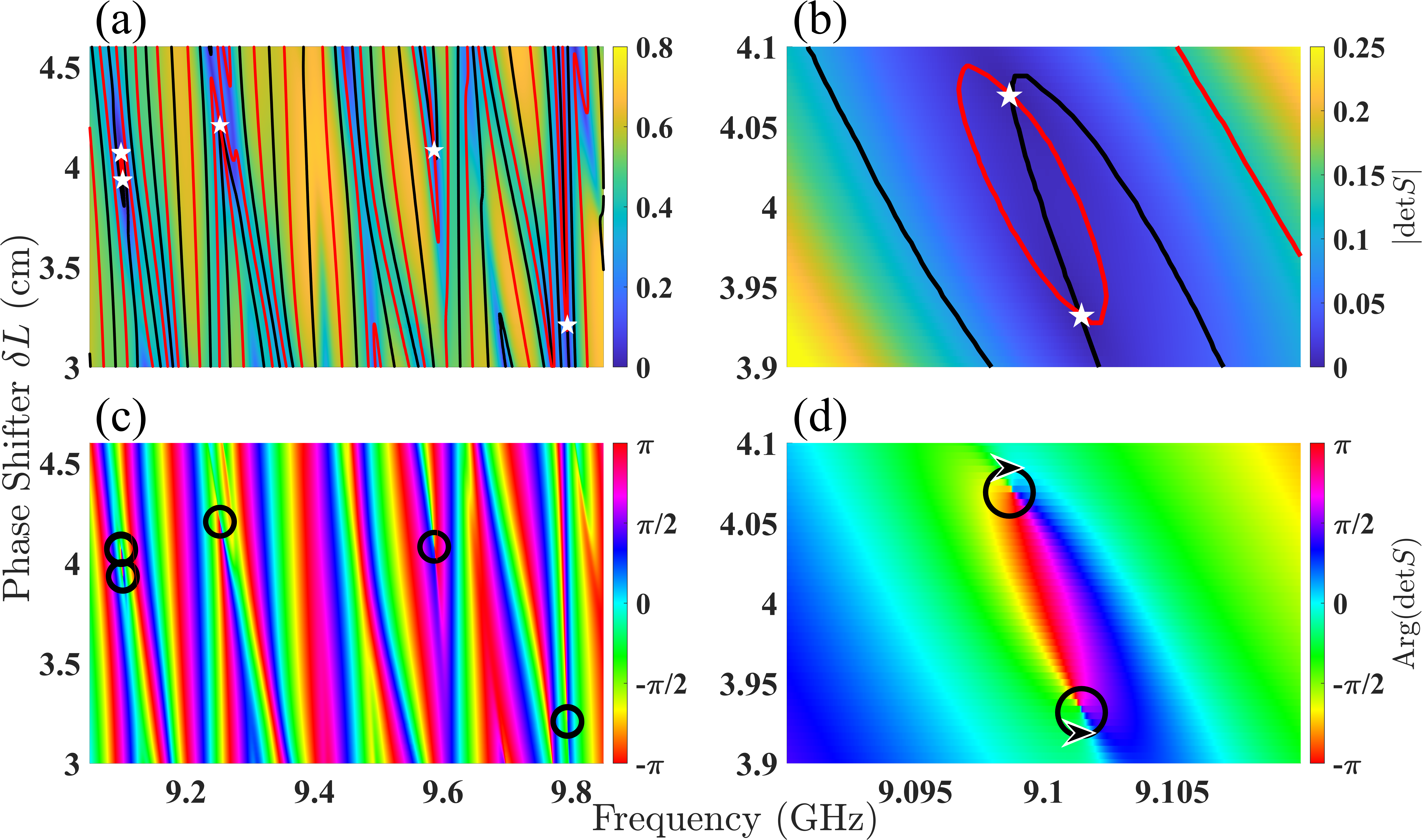}}
\caption{Speckle pattern of scattering singularities in $d=2$ parameter space. The experimental S-matrix data comes from a tetrahedral microwave graph. (a-b) Magnitude of $\text{det}S$, red lines mark locations of $\text{Re}[\text{det}S]=0$ while black lines mark locations of $\text{Im}[\text{det}S]=0$. White symbols highlight intersections which correspond to $\text{det}S=0+i0$ which are CPAs. (c-d) Phase of $\text{det}S$, black circles show locations of phase windings which align exactly with the white symbols in (a-b). Arrows on the circles in (d) show the two nearby CPAs have opposite winding numbers.}
\label{Fig1}
\end{figure*}

Fig.~\ref{Fig1} is a plot of $\text{det}S$ in a $d=2$ parameter space, obtained directly from experimental data of a wave scattering system to be described below. The red (black) lines in panels (a) and (b) correspond to the topologically stable curves $\text{Re}[\text{det}S]=0$ ($\text{Im}[\text{det}S]=0$). These curves either form closed loops or extend beyond the measured parameter space. Their intersections at $\text{det}S=0+i0$ (CPA) are marked by white symbols. There are many CPA points in panel (a), while panel (b) shows a close up view of a pair of CPA points. In panels (c) and (d), the phase of $\text{det}S$ is plotted with a cyclic color map. The location of phase singularities marked by the black circles align exactly with the white symbols in (a) and (b). As stated earlier, a similar picture could be made by highlighting locations of $\text{Re}[\text{det}S]=a$ and $\text{Im}[\text{det}S]=b$ as well as plotting the phase of $\text{det}S-(a+ib)$.

% \begin{figure}[bth]
% \includegraphics[width=0.48\textwidth]{Fig1_v3.png}
% \caption{Speckle pattern of singularities in $d=2$ parameter space. The experimental S-matrix data comes from a tetrahedral microwave graph. (a-b) Magnitude of $\text{det}S$, red lines mark locations of $\text{Re}[\text{det}S]=0$ while black lines mark locations of $\text{Im}[\text{det}S]=0$. White symbols highlight intersections which correspond to $\text{det}S=0+i0$ which are CPAs. (c-d) Phase of $\text{det}S$, black circles show locations of phase windings which align exactly with the white symbols in (a-b). Arrows on the circles in (d) show the two nearby CPAs have opposite winding number.}
% \label{Fig1}
% \end{figure}

Another singularity of interest is an Exceptional Point degeneracy of order $P$ (EP-P) where $P\geq2$ eigenvalues and eigenvectors of a non-Hermitian operator become degenerate \cite{Berry2004,El-Ganainy18,Alu19,Özdemir19,Ashida20,Ding2022}. Similar to CPAs, EPs have been explored in many fields such as optics \cite{PhysRevLett.106.213901,PhysRevLett.112.203901,Hodaei2017,Chen2017}, microwave cavities \cite{PhysRevLett.86.787,PhysRevE.69.056216,PhysRevLett.106.150403}, exciton-polaritons \cite{Gao2015}, acoustics \cite{Shi2016}, waveguides \cite{Doppler2016}, electrical circuits \cite{Stehmann_2004,Suntharalingam2023}, etc. In this paper we specifically consider EPs of the $S$-matrix, which \rev{in general} are unrelated to the Hamiltonian EPs, except for specially contrived cases such as in Ref.~\cite{xia2025}. The EPs of the $S$-matrix are the same as those for the impedance $Z$, admittance $Y$, and Wigner reaction $K$ matrices \cite{erb2025}. EP-2s are topologically protected as they are the zeros of the complex scalar field $\lambda_j-\lambda_k$ where $\lambda_k$ is the $k$th eigenvalue of an operator $\mathcal{H_S}$. Exceptional Points of order higher than $2$ are not topologically protected, as will be discussed later.

\rev{Due to the eigenvalue repulsion, the raw eigenvalue difference $\lambda_j-\lambda_k$ can actually be a poor quantity to use for identifying Exceptional Points. We instead use} the eigenvector coalescence measure $|C_{jk}|$ where:
\begin{equation}
    |C_{jk}|  = \frac{|\langle R_j|R_k\rangle|}{|R_j||R_k|} \nonumber
\end{equation}
is the normalized inner-product of two right eigenvectors of the scattering matrix such that $S \ket{R_j} = \lambda_j \ket{R_j}$\cite{erb2025}. The quantity $|C_{jk}|$ is bounded between $0$ ($\ket{R_j}$ and $\ket{R_k}$ are orthogonal) and $1$ ($\ket{R_j}$ and $\ket{R_k}$ are degenerate, namely an EP-2). For a unitary scattering matrix \rev{which describes a lossless system}, the eigenvectors are always orthogonal, so there are no Exceptional Points, only Diabolic Points (DP) \cite{Chen2025}. A subunitary scattering matrix can have orthogonal eigenvectors, but only as a special case, which makes the eigenvector orthogonality condition also of interest. We find that for a system with Lorentz reciprocity, two-eigenvector orthogonality has the dimension of curves (similar to the red and black lines in Fig.~\ref{Fig1}), but if reciprocity does not hold then the orthogonality takes the dimensionality of points \cite{erb2025}. Going forward, only the coalescence of two eigenvectors will be considered in this paper, and the $jk$ notation will be suppressed. The eigenvector coalescence is related to the Petermann factor $K=\frac{1}{1-|C|^2}$ \rev{and the phase rigidity or condition number $r=\sqrt{1-|C|^2}$} \cite{Alireza24,Tuxbury2022,Wiersig23,Kullig2025}.

Other simple singularities include matrix-element zeros of the $S$-matrix. The complex zeros of $S_{xx}$ or $S_{xy}$ are known as Reflectionless Scattering Modes (RSMs) and Transmissionless Scattering Modes (TSMs) \cite{Sol2023,Davy2023,Jiang2024,Faul2025,Kang2021,Asano2016,Huang2022,Sweeney2020,Stone2021,Genack2024}. In a system with broken Lorentz reciprocity, it is possible to have $S_{xy}=0$ or $S_{yx}=0$ independently of each other, and when only one equals zero there is Uni-Directional Transmission. Further we can define $\delta R_{xy}:=S_{xx}-S_{yy}$ and $\delta T_{xy}:=S_{xy}-S_{yx}$ as the reflection and transmission differences between channels $x$ and $y$. If $\delta R_{xy}=0+i0$, then the two channels have symmetric reflections, which is an interesting singularity when it occurs in complex scattering systems without any geometric symmetries. Similarly if $\delta T_{xy}=0+i0$, then transmission between the two channels is reciprocal and \rev{we find that} this singularity is present in systems with broken Lorentz reciprocity \cite{erb2025}. 

As stated before, the divergence of complex time delays is associated with scattering singularities: Wigner-Smith $\tau_{WS}= \frac{-i}{M}\frac{\partial}{\partial E}\text{log}[\text{det}S]$ with CPA, Reflection $\tau_{xx}= -i\frac{\partial}{\partial E}\text{log}[S_{xx}]$ with RSM, Transmission $\tau_{xy}= -i\frac{\partial}{\partial E}\text{log}[S_{xy}]$ with TSM, etc. Since complex time delay itself can also take the value $0+i0$, there are additional scattering singularities that can be identified with complex time delay. To see why the zero of complex time delay is interesting, we turn to recent work on interpreting complex time delay in the frequency domain as the time and center frequency shift of a pulse in the time domain \cite{Asano2016,giovannelli2025}. This implies that the zero of complex time delay corresponds to a minimal distortion point of the system, at which signals are only attenuated (or amplified), but not otherwise distorted after propagation.

There are a myriad of other scattering singularities that we could discuss, and a more expansive listing is provided in Appendix \ref{App_listsings}. 

\subsection{Higher Order Singularities} \label{Sub_HighSing}
All the singularities given so far are topologically stable, being the zero of a single complex scalar field $\mathcal{S}$. In experimental systems, we can use embedded \rev{tunable} perturbers to create collisions of fundamental singularities of different fields ($\mathcal{S}$ and $\mathcal{S}'$) at real frequencies in a given parameter space. These collisions are not topologically protected and a small fluctuation in the system can move the singularities apart. Despite their instability, such ``higher order'' singularities can have novel properties and further applications, and so should be of great interest.

One example of a higher order singularity is the $S$-matrix CPA-EP demonstrated in Ref.~\cite{erb2025} to turn a complex, irregular tetrahedral graph into a robust 50:50 In-Phase/Quadrature microwave splitter. CPA-EPs have also been found in acoustic systems \cite{xia2025}, and in the spectrum of the Hamiltonian \cite{Sweeney2019,Wang2021}.  There are also higher order $S$-matrix EP-Ps which are intersections of exactly P$-1$ unique EP-2s. Because the eigenvalue detuning around an EP has a $\frac{1}{P}$ power law, it shows extreme sensitivity to small perturbations, leading to proposals of EP sensors \cite{PhysRevLett.112.203901,Hodaei2017,Chen2017,Kononchuk22}. RSMs and TSMs can also be made degenerate, for example if both $S_{xx}$ and $S_{yy}$ equal zero simultaneously then the system has an RSM-2, sometimes called a RSM or RL EP \cite{Ferise2022} because the eigenvectors of a reflectionless operator become degenerate. RSM-2s have been used to create broadband perfect transmission \cite{zhumabay2025}. Reciprocal systems have fundamental TSM-2s and can only have TSM-Ps of \textit{even} order P.

 Another set of interesting higher order singularities are $S$-matrix row/column-zeros. The row-zero condition $\sum_y^M|S_{xy}|=0$ describes a situation in which no power leaves the system through port $x$, no matter which port the initial power was injected into the system through. The column-zero condition $\sum_x^M|S_{xy}|=0$ describes a situation in which all the power injected into the system through port $y$ is completely absorbed. Of note is that in a reciprocal system, these conditions are equivalent and both phenomena happen simultaneously. When both row and column $j$ are completely zero, the $S$-matrix is essentially reduced in dimensionality as one channel appears to not participate in the scattering interaction, acting solely as an absorber.

Again, there are many ways to combine fundamental singularities into higher order singularities, but none of these combinations are topologically stable and the more fundamental singularities included, the less probable such events become \rev{in a statistical ensemble}.

\subsection{Paper Overview}

This paper is structured as follows. Section \ref{Sec_Exp} describes the experimental systems considered and their parametric degrees of freedom, as well as the measurements. In section \ref{Sec_Results}, the experimental data is presented and discussed in three subsections: \ref{Sub_FunStats} has the statistics of fundamental singularities, \ref{Sub_SingAbs} presents the absorption dependence of singularity abundance, and \ref{Sub_HighStats} contains the statistics of higher order singularities, which are composite of multiple fundamental singularities. There are also appendices \ref{App_Proof}-\ref{App_listsings} which document details and additional data supporting the claims made in the text.

Although the experimental data presented in this paper all comes from microwave systems, the agreement shown with Random Matrix Theory (RMT) formulation of the scattering matrix indicates that the conclusions are generic for all wave scattering systems. We suggest that this is not the limit, and the ``superuniversal'' statistics found in these systems extend to many other fields of physics where similar analysis can be performed\rev{, such as the statistics of transmission zeros in optics or singularities of quantum weak measurements as seen in Ref.~\cite{Solli2004}}.

\section{Experiment}\label{Sec_Exp}
\begin{figure}[bht]
\includegraphics[width=0.48\textwidth]{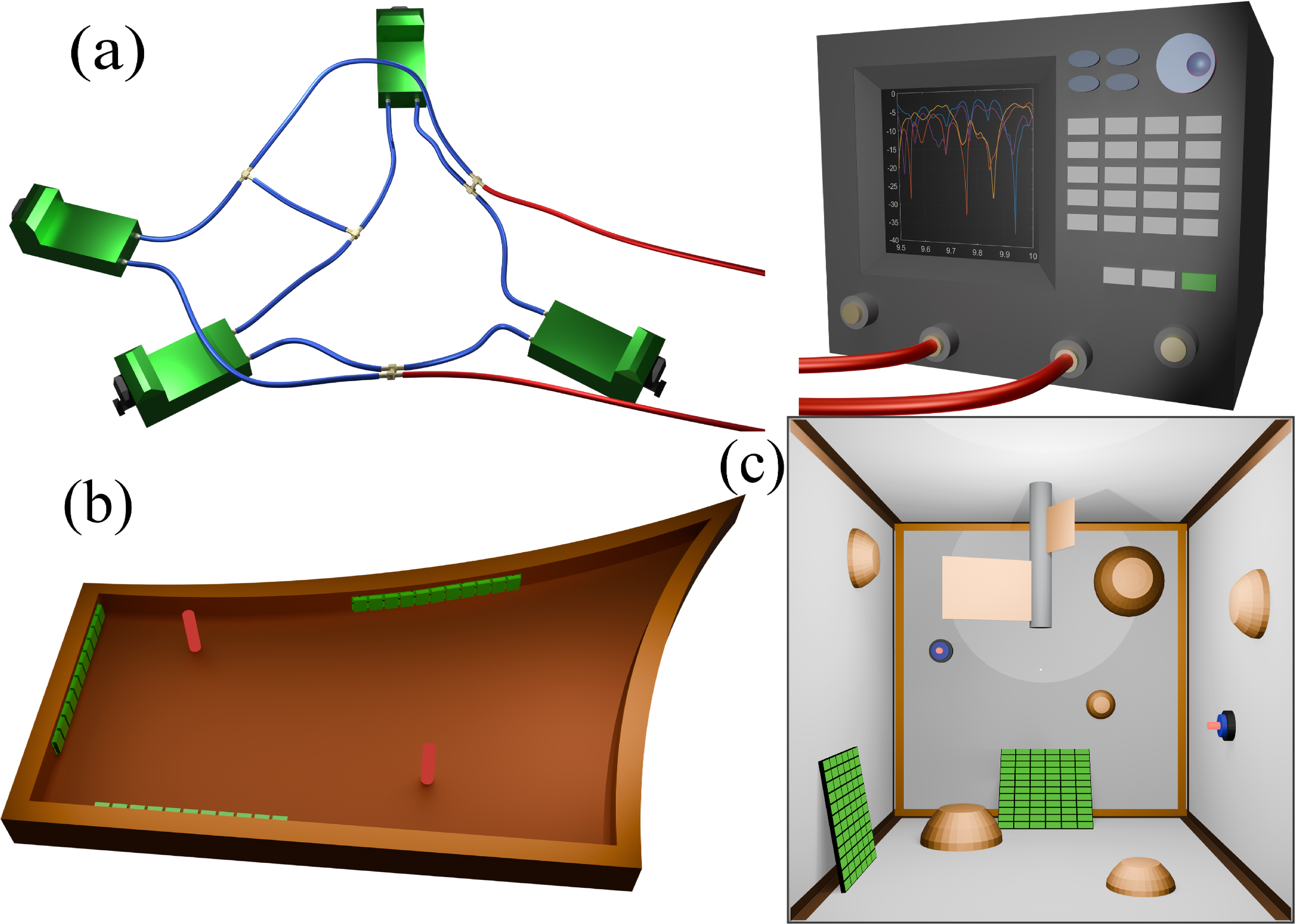}
\caption{Schematics of experimental systems. (a) A tetrahedral microwave graph ($\mathcal{D}=1$), (b) a ray-chaotic quarter bowtie billiard ($\mathcal{D}=2$), and (c) a three dimensional cavity with various symmetry breaking elements ($\mathcal{D}=3$). Scattering channels connected to the network analyzer (top right) are marked in red, and embedded perturbers are marked in green.}
\label{Fig2}
\end{figure}

%\textit{Experiment}.---
We measure the $M\times M$ $S$-matrix of various complex microwave scattering systems with calibrated Keysight PNA-X N5242A and PNA-X N5242B microwave vector network analyzers. By complex we mean that the systems are excited with waves whose wavelengths are small compared to the system size, and interference effects for waves following different ray trajectories are extremely sensitive to details of the system configuration. The systems in question are depicted in Fig.~\ref{Fig2}: (a) a tetrahedral graph ($\mathcal{D}=1$), (b) a ray-chaotic quarter bowtie billiard ($\mathcal{D}=2$), and (c) a three dimensional cavity with various symmetry breaking elements ($\mathcal{D}=3$), where $\mathcal{D}$ is the wave propagation dimension. For clarity in the schematic, the $M=2$ scattering channels connected to the \rev{network analyzer} are shown in red, and the embedded tunable perturbers are shown in green. In the graph, the tunable perturbers take the form of voltage controlled mechanical phase shifters on four of the six bonds \cite{Rehemanjiang2018,Dietz2017,Che2021,Che2022,Lawniczak2023}. These phase shifters act as variable-length cables $L+\delta L$, allowing us to change the interference conditions at the nodes. Mechanical phase shifters are superior to digital phase shifters for this purpose, as digital phase shifters typically have much higher insertion loss, whereas the mechanical phase shifters have loss comparable to the coaxial cables that make up the rest of the graph, and have very precise step size resolution. In the billiard and three dimensional cavity there are globally-biased, voltage controlled varactor-loaded metasurfaces \cite{Chen2016,Elsawy2023,Frazier2022,Sleasman2023,Erb2024}. The metasurfaces give us control over the  amplitude and phase of reflected waves, and in the reverberant scattering environments most waves will interact with these metasurfaces multiple times, allowing for impactful control over the scattering properties of the systems.

The tunable perturbers make two types of measurements possible. One is a continuous parameter sweep like the kind shown in Fig.~\ref{Fig1} where typically a single perturber is tuned with very fine steps between frequency scans, while the rest of the system parameters are held fixed. This kind of measurement reveals the phase winding around singularities, and through the use of a second perturber the dynamics of the singularities can be visualized \cite{Erb2024,erb2025}. The other kind of measurement is a statistical ensemble, where each perturber is tuned simultaneously to very different values between frequency scans. This ensures that the difference between each realization in the ensemble is statistically significant. In other words, the first kind of measurement gives the scattering matrix in a closed area ($d=2$) or volume ($d=3$) of parameter space, while the second kind of measurement provides a reasonably ergodic sampling of a higher dimensional parameter space ($d\geq4$). All the statistical results discussed in this paper will be derived from the second kind of measurement.

We consider four global system parameters as being important and hold these fixed across realizations of a given ensemble. These are: the dimension for wave propagation $\mathcal{D}$ ($\mathcal{D}\in \{1,2,3\}$), number of scattering channels $M$ ($M\in \{1,\dots,\infty\}$, although our \rev{network analyzer} has a limit of $M\leq4$), Dyson class $\beta$ ($\beta\in \{1,2\}$), and  uniform \rev{cavity} dissipation $\eta$ ($\eta\in \{0,\dots,\infty\}$). The Lorentz reciprocity of a system can be broken, changing $\beta$ from 1 to 2, by adding a magnetized ferrite in the propagation path of the microwaves \cite{So1995,Stoffregen1995,Dietz2007,Dietz2009,Lawniczak2009,Lawniczak2010,Lawniczak2011,Bittner2014,Castaneda2022}. \rev{We approximate the cavity loss to be effectively homogeneous in all our experimental systems. In the $\mathcal{D}=1$ graph the absorption is due to dielectric loss and conductor resistivity in the coaxial cables and phase shifters, which are nominally uniform \cite{Lei2022}. In the $\mathcal{D}=2$ billiard, the majority of the loss is caused by induced currents on the top and bottom plates which have uniform resistivity. Finally, the $\mathcal{D}=3$ cavity is highly reverberant, meaning the waves interact with the lossy boundaries many times.}

Note that $\eta$ is a dimensionless quantity given by $\eta := \tau_H\tilde{\eta}$ where $\tilde{\eta}$ is an \rev{internal dissipation} rate proportional to the average mode bandwidth \cite{Buttiker1986,Brouwer1996,Kuhl2013}. The Heisenberg time  $\tau_{H} = \frac{2\pi}{\Delta}$ can be related to the mean \rev{delay} time of waves in a scattering environment, where $\Delta$ is the mean mode spacing in frequency units. Hence, the \rev{cavity} absorption strength $\eta$ is inherently tied to our system dimension $\mathcal{D}$, as the larger systems have longer \rev{delay} times which cause the waves to be more strongly attenuated. Since both $\tilde{\eta}$ and $\tau_H$ (for $\mathcal{D}>1$) are frequency dependent \cite{Lei2022}, by measuring \rev{the same system} in different frequency bands we can systematically vary the value of $\eta$.  Distributing absorbers in a cavity \cite{Hemmady2005,Hemmady2005b,Hemmady2006,Hemmady2006c} or attenuators on the bonds of a graph \cite{Hul2005,Lawniczak2023,Lawniczak2019} is another way to change $\eta$, though these methods can only increase the loss. One can also employ cryogenic cooling to decrease $\eta$ \cite{Graf92,Hemmady2006,Dietz2015,Xiao2018,Ma2023}.  For this paper we have prepared systems with $\eta$ varying from $1.8$ to $145$, which covers well what has been considered the low, moderate, and high absorption strength regimes \cite{Sanchez2003,Hul2004,Hemmady2005,Fyodorov2004,Savin2005,Lawniczak2008,Lawniczak2019}. \rev{Table~\ref{ENS_Table} in Appendix~\ref{App_ENS} has the values of the parameters and number of realizations in each measured ensemble.}

The $S$-matrix measured by a \rev{network analyzer} is dependent on how well the scattering channels are coupled to the enclosure. Our systems are designed to be well coupled, but the coupling is still imperfect and frequency dependent. We remove these non-universal effects from ensembles through application of the Random Coupling Model (RCM) normalization process, which results in $S$-matrices with perfect coupling \cite{Hemmady2005,Zheng20052Port,Zheng2006,Hart2009,Yeh2010, Lawniczak2012,Hemmady2012,Lee2013,Yeh2013,Gradoni2014,Auregan16,Dietz2017,Fu2017}. Coupling effects cannot be removed from parametric studies like in Fig.~\ref{Fig1} because RCM requires statistical information. The effects of coupling on scattering singularity statistics are not addressed in this paper and will be explored in subsequent work. For more details on the experimental systems, such as more information on the tunable pertubers or how an ensemble value of $\eta$ is calculated, refer to Refs.~\cite{shaibe2025} and \cite{erb2025}.

\section{Data and Discussion} \label{Sec_Results}

In this section we present statistical results of scattering singularities from both experiment and RMT numerics. Note that every scattering matrix included is perfectly coupled, see Section \ref{Sec_Exp}. We consider PDFs of functions that diverge at singularities for two reasons: (1) divergent functions are more advantageous for finding singularities, as compared to complex linear functions, because they require only one degree of freedom and have a more telling character when singularities occur, and (2) the PDF of a divergent function is dominated by an infinitely long tail which corresponds to the presence of singularities. This means most of the PDF is of importance when considering singularities, rather than just a small region around the singularity value. In particular, for every fundamental singularity, it is possible to define a time delay or some generalized delay that diverges at the singularity, as discussed in Appendix~\ref{App_CTD}. In subsection \ref{Sub_FunStats} we present the statistics of topologically protected points (fundamental singularities) and curves in two dimensional parameter space. Subsection \ref{Sub_SingAbs} focuses on how absorption affects singularity abundance in \rev{lossy} systems because absorption appears to be the most important determining parameter. Lastly, in subsection \ref{Sub_HighStats} we discuss higher order singularities which show system-specific statistics, in contrast to fundamental singularities.

\subsection{Fundamental Singularities - Statistics} \label{Sub_FunStats}

\begin{figure*}[bth]
\includegraphics[width=\textwidth]{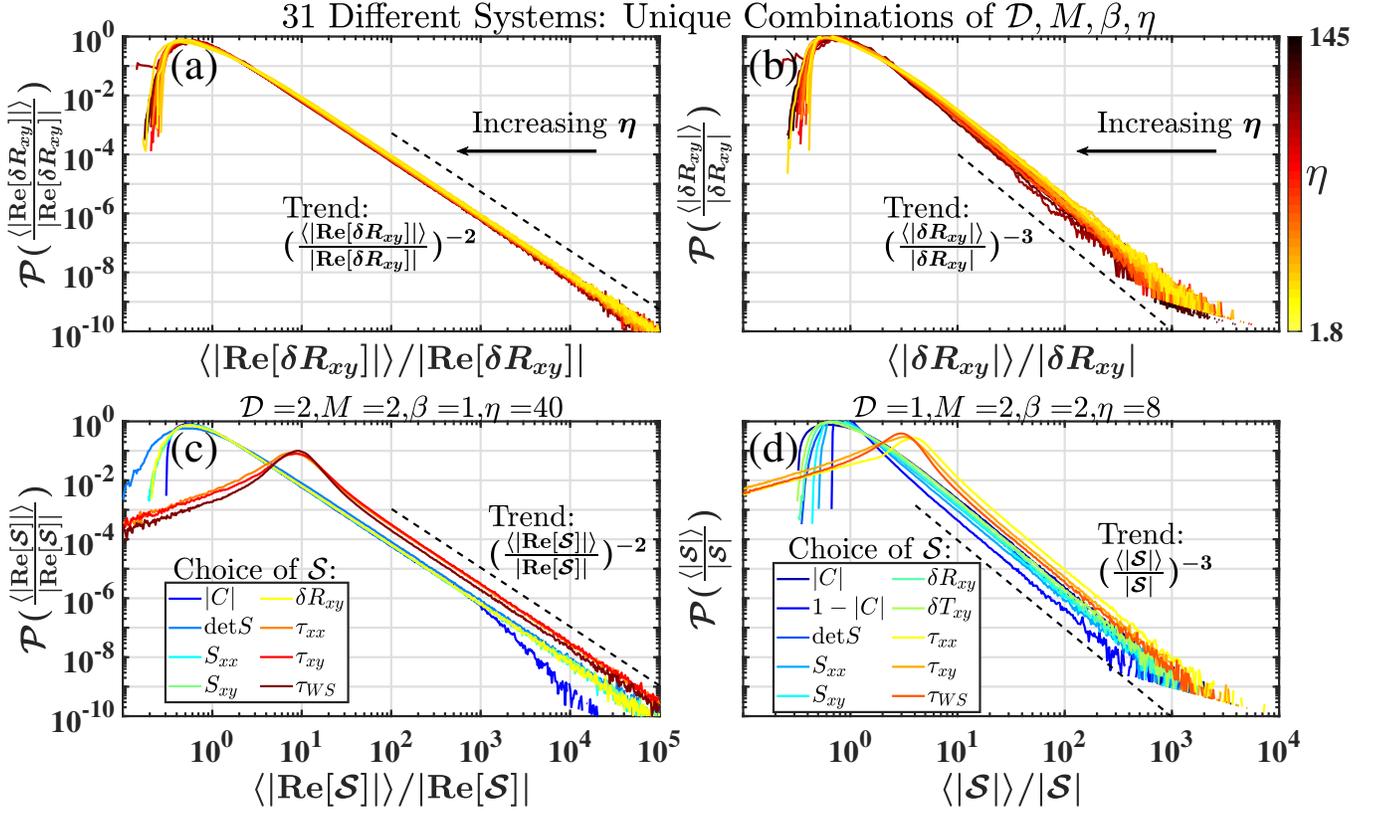}
\caption{(a-b) PDFs of $\frac{\langle |\text{Re}[\delta R_{xy}]| \rangle}{|\text{Re}[\delta R_{xy}]|}$ and $\frac{\langle |\delta R_{xy}| \rangle}{|\delta R_{xy}|}$ from 31 ensembles of experimental scattering systems with different values for the four parameters $\mathcal{D},M,\beta,\eta$ \rev{(see Appendix \ref{App_ENS} for an enumeration of the ensembles)}.  Color gradient of curves corresponds to $\eta$ (not uniformly spaced) of the ensembles. Dashed black lines characterize \rev{the} power law of \rev{the} tail behavior. (c-d) PDFs of $\frac{\langle|\text{Re}[\mathcal{S}]| \rangle}{|\text{Re}[\mathcal{S}]|}$ and $\frac{\langle |\mathcal{S}|\rangle}{|\mathcal{S}|}$ for various functions $\mathcal{S}$ from two arbitrary ensembles. Dashed black lines characterize \rev{the} power law of \rev{the} tail behavior.}
\label{Fig3}
\end{figure*}

In Fig.~\ref{Fig3}(a-b), we show the logarithmic PDFs of $\frac{\langle |\text{Re}[\delta R_{xy}]| \rangle}{|\text{Re}[\delta R_{xy}]|}$ and $\frac{\langle|\delta R_{xy}| \rangle}{|\delta R_{xy}|}$ from 31 ensembles of experimental scattering systems with different values for the four parameters $\mathcal{D},M,\beta,\eta$ \rev{(see Appendix \ref{App_ENS} for the details and Table~\ref{ENS_Table} which has the parameter values for each ensemble presented)}. \rev{H}ere\rev{,} $\delta R_{xy}=S_{xx}-S_{yy}$ is the reflection difference between ports $x$ and $y$, and $\langle|\delta R_{xy}| \rangle$ is the ensemble mean of $|\delta R_{xy}|$. We normalize by the ensemble mean value to bring the PDFs together, as otherwise the PDFs from different systems can be distantly shifted horizontally from each other, making it difficult to compare the tails. We find that $\mathcal{P}(\frac{\langle |\text{Re}[\delta R_{xy}]|\rangle}{|\text{Re}[\delta R_{xy}|]})$ has a universal $-2$ power law tail that holds for all systems, represented by the dashed black line in panel (a), since $\frac{\langle |\text{Re}[\delta R_{xy}]| \rangle}{|\text{Re}[\delta R_{xy}]|}$ \rev{has a simple pole} at $\text{Re}[\delta R_{xy}]=0$, a constraint on a single degree of freedom. Similarly, $\frac{\langle |\delta R_{xy}|\rangle }{|\delta R_{xy}|}$ \rev{has a simple pole} at $\delta R_{xy}=0+i0$ which are points, specifically symmetric reflection points, and we find that $\mathcal{P}(\frac{\langle |\delta R_{xy}| \rangle}{|\delta R_{xy}|})$ has a universal $-3$ power law represented by the dashed line in panel (b).

Three arbitrary choices were made in constructing Fig.~\ref{Fig3}(a-b). First, the choice of complex scalar function to investigate. There is nothing special about the reflection difference $\delta R_{xy}$, any function with singularities will produce the same statistical results. Second, in panel (a) we could have used the imaginary part of $\delta R_{xy}$ (as opposed to the real part) and the PDF tail behavior would have been the same; there is no particular reason to chose one over the other. Finally, plotting the PDFs of the magnitude, without regard to sign. This choice was made for simplicity as plotting both positive and negative values on a logarithmic scale is non-trivial. That was done, however, in Refs.~\cite{Lei2021Stats,shaibe2025} for the case of complex time delay, and there the same tails were seen on both the positive and negative sides of the distribution.

In Fig.~\ref{Fig3}(c), we show the logarithmic PDFs of eight different functions which \rev{have simple poles} at topologically protected curves: (i) $\frac{\langle|C|\rangle}{|C|}$, (ii) $\frac{\langle|\text{Re[det}S]|\rangle}{|\text{Re[det}S]|}$, (iii) $\frac{\langle|\text{Re}[S_{xx}]|\rangle}{|\text{Re}[S_{xx}]|}$, (iv) $\frac{\langle|\text{Re}[S_{xy}]|\rangle}{|\text{Re}[S_{xy}]|}$, (v) $\frac{\langle|\text{Re}[\delta R_{xy}]|\rangle}{|\text{Re}[\delta R_{xy}]|}$, (vi) $\frac{\langle|\text{Re}[\tau_{xx}]|\rangle}{|\text{Re}[\tau_{xx}]|}$, (vii) $\frac{\langle|\text{Re}[\tau_{xy}]|\rangle}{|\text{Re}[\tau_{xy}]|}$, and (viii) $\frac{\langle|\text{Re}[\tau_{WS}]|\rangle}{|\text{Re}[\tau_{WS}]|}$. Just as in panel (a), the same thing could be done using the PDFs of the imaginary parts of these functions rather than the real, except for $\frac{\langle|C|\rangle}{|C|}$ as $|C|$ is already a real number.  All the data in panel (c) comes from one ensemble, a reciprocal two channel billiard ($\beta=1,M=2,\mathcal{D}=2$) with uniform absorption of $\eta=40$. Because the functions \rev{have simple poles} at curves, the PDFs all have $-2$ power laws as represented by the dashed black curve. The deepest blue curve corresponding to $\mathcal{P}(\frac{\langle |C| \rangle}{|C|})$ has a \rev{downturn} and transitions to a $-3$ power law. This can be explained by the fact that our experimental systems always have some small degree of non-reciprocity which is impossible to eliminate. Further, the \rev{network analyzer} determines $S_{xy}$  and $S_{yx}$ independently and does not have arbitrary precision. For most purposes, this is a negligible effect but in non-reciprocal systems two-eigenvector orthogonality takes the dimension of points \rev{because it requires two conditions to be simultaneously satisfied (see Appendix C of Ref.~\cite{erb2025})}, which induces the $-3$ power law. In RMT simulations that are perfectly reciprocal, this transition in the power law of the PDF tail from $-2$ to $-3$ is not seen (see Appendix \ref{App_RMT}).

Panel (d) of Fig.~\ref{Fig3} has the logarithmic PDFs of ten different functions which \rev{have simple poles} at topologically protected points: (i) $\frac{\langle|C|\rangle}{|C|}$ diverges at Orthogonality Points, (ii) $\frac{1-\langle|C|\rangle}{1-|C|}$ diverges at Exceptional Points, (iii) $\frac{\langle|\text{det}S|\rangle}{|\text{det}S|}$ diverges at Coherent Perfect Absortion, (iv) $\frac{\langle|S_{xx}|\rangle}{|S_{xx}|}$ diverges at Reflectionless Scattering Modes, (v) $\frac{\langle|S_{xy}|\rangle}{|S_{xy}|}$ diverges at Transmissionless Scattering Modes, (vi) $\frac{\langle|\delta R_{xy}|\rangle}{|\delta R_{xy}|}$ diverges at Symmetric Reflection Points, (vii) $\frac{\langle|\delta T_{xy}|\rangle}{|\delta T_{xy}|}$ diverges at Reciprocal Transmission Points, (viii) $\frac{\langle|\tau_{xx}|\rangle}{|\tau_{xx}|}$ diverges at Complex Reflection Time Delay zeros, (ix) $\frac{\langle|\tau_{xy}|\rangle}{|\tau_{xy}|}$ diverges at Complex Transmission Time Delay zeros, and (x) $\frac{\langle|\tau_{WS}|\rangle}{|\tau_{WS}|}$  diverges at Complex Wigner-Smith Time Delay zeros. All the data in this panel comes from an ensemble measurement of a non-reciprocal two channel graph ($\beta=2,M=2,\mathcal{D}=1$) with uniform absorption of $\eta=8$. Eigenvector orthogonality has dimension of $d-2$ in non-reciprocal systems, but dimension of $d-1$ in reciprocal systems, which is why $\mathcal{P}(\frac{\langle |C| \rangle}{|C|})$ appears in both panels (c) and (d). Unlike panel (c), there is no deviation from the expected $-3$ power law behavior for the tails in panel (d). 

Between panels (a) and (c), it should become clear that the PDF of every function in every system that diverges \rev{with a simple pole} at a topologically protected curve has a $-2$ power law. Similarly, panels (b) and (d) convey that the PDF of every function in every system that diverges \rev{with a simple pole} at a topologically protected point has a $-3$ power law. A very simple mathematical argument for why the universal power laws take these values is given in Appendix \ref{App_Proof}. A figure showing the power laws of PDFs calculated from $S$-matrices constructed with RMT numerics is available in Appendix \ref{App_RMT}. The fact that RMT numerics returns statistics for scattering singularities which is identical to our empirical results from microwave systems reveals that these statistics are not specific to any \rev{particular} kind of wave scattering. Rather, these statistical results describe all generic scattering systems. 

\rev{Notably, in Fig.~\ref{Fig3} we do not present the PDFs of the inverse eigenvalue difference $\frac{1}{{|\lambda_j-\lambda_k|}}$ nor the inverse eigenvalue condition number $r^{-1}$. This is because neither of these functions have a simple pole at a singularity. Because of the eigenvalue repulsion mentioned before, the eigenvalue difference $|\lambda_j-\lambda_k|$ goes to zero non-linearly, hence $\frac{1}{{|\lambda_j-\lambda_k|}}$ is not a simple pole. Similarly, $r^{-1}$ actually has a branch point of the form $\frac{1}{\sqrt{(1-|C|)}}$ at an EP-2. Since they do not satisfy the condition of having a simple pole divergence only at a singularity, the PDFs of $\frac{1}{{|\lambda_j-\lambda_k|}}$ and $r^{-1}$ do not have $-3$ power laws. We provide some examples of PDFs of functions that diverge at singularities but not as simple poles in Appendix \ref{App_Tails}.}

Since a distribution with a $-3$ power law has an undefined variance, a generic non-Hermitian scattering system must have every kind of scattering singularity allowed somewhere in its parameter space. However, it may not necessarily be accessible depending on the tunability of the experimental perturbers. Note also that since the $-3$ power law tail is ``superuniversal'', the PDF of a function that diverges at a singularity is the perfect platform for investigating which global system parameters have the greatest effect on singularity abundance and density. It is also worth considering what the limitations of the ``superuniversality'' are, and whether there are ways to break it, for example by having imperfect coupling which was seen to disrupt the $-3$ power laws from some but not all complex time delay PDFs in Ref.~\cite{shaibe2025}.

In panels (a-b) of Fig.~\ref{Fig3}, the color gradient corresponds to the absorption strength $\eta$, and there is a systematic, though small, vertical offset of the PDFs based on the ensemble value of $\eta$. This leads to different abundances of extreme values of $\mathcal{S}$.  It is clear that the degree of loss is the strongest predictor of singularity density in parameter space. For any given quantity, the PDFs from ensembles with higher loss turn over to the power law tail at smaller values of the quantity and therefore have a lower overall probability at the extreme values associated with the scattering singularities. This agrees with the finding in \cite{shaibe2025}, that the time delay PDFs from ensembles with larger $\eta$ had the power law tails onset earlier, resulting in smaller probabilities at extreme values, hence fewer singularities in systems with more uniform absorption. \rev{Note, however, that this is only true for systems with sufficient absorption. We do not have experimental data from an ensemble with $\eta<1.8$, but since certain singularities are impossible in lossless systems (consider CPA which requires $\text{det}S=0+i0$, but if $\eta=0$ then $|\text{det}S|=1$) we expect there must be some range of $\eta$ where increasing loss increases the abundance of singularities.}

\subsection{Singularity Dependence on Absorption} \label{Sub_SingAbs}

\begin{figure}[bht]
\includegraphics[width=0.48\textwidth]{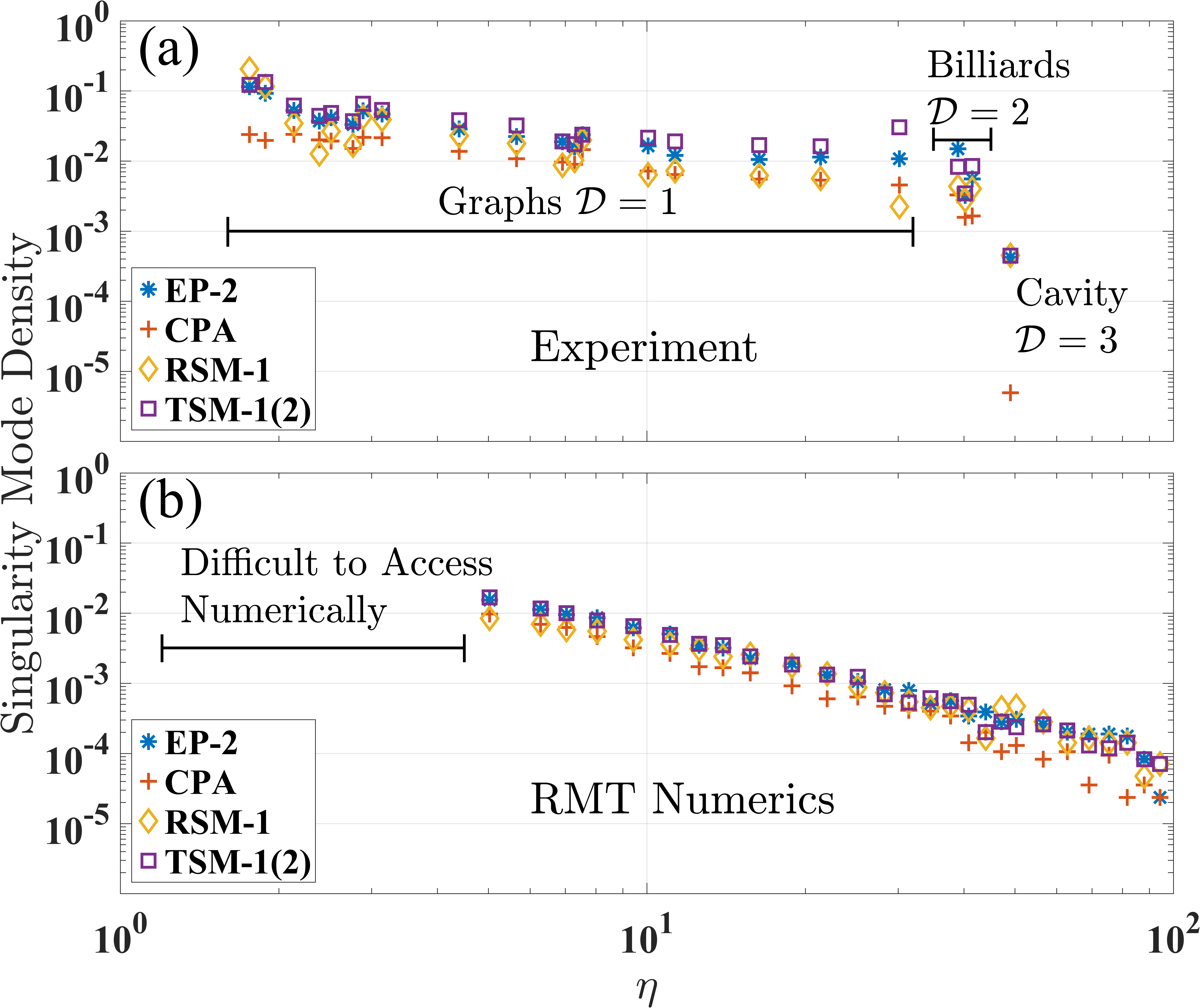}
\caption{Singularity density as a function of \rev{uniform} loss in (a) experimental ensembles and (b) RMT simulation. The overall trend of reduced singularity density with increasing loss \rev{in this finite range of $\eta$} is easier to see in (b) as the experimental systems have more differences than just $\eta$, which cannot all be normalized out.}
\label{Fig4}
\end{figure}

It might seem counterintuitive that singularity abundance would decrease with increasing absorption since singularities are zeros of the complex wave fields and absorption should suppress the fluctuation magnitude of the waves. However, singularities, as stated before, are primarily the result of interference effects of superpositions of random waves. In \rev{systems with large} dissipation, highly attenuated waves will not experience much destructive interference because the \rev{wave excitations} become more homogeneous and less random.

In Fig.~\ref{Fig4}(a)  we show the singularity mode density, defined as the number of singularities per mode in the ensemble, of the fundamental singularities EP-2, CPA, RSM-1, and TSM-1s (or equivalently TSM-2s for reciprocal $\beta=1$ systems) found in our experimental ensembles as a function of $\eta$. As discussed in Sec.~\ref{Sec_Exp}, the mean mode spacing $\Delta$ is system size and frequency dependent, so $\eta$ is correlated to $\mathcal{D}$ in our experimental systems. Also, the techniques used to create ensembles are different for each $\mathcal{D}$ so despite the great effort made to create diverse ensembles, there is some systematic dependence for how `ergodically' we can explore the phase space of a given scattering environment based on what tunable parameters we have available.  However, we still see a trend of lower singularity density with increasing $\eta$, agreeing with the complex time delay PDFs in Refs.~\cite{Lei2021Stats,shaibe2025} that \rev{downturn into the $-3$ tail} at smaller time delay values for systems with larger $\eta$. We performed RMT simulations using a Gaussian Orthogonal Ensemble (GOE) of 600 random matrices of size $10^5\times10^5$ to generate scattering matrices following the procedure detailed in Appendix \ref{App_RMT}, Ref.~\cite{Lee2013} and Appendix A of Ref.~\cite{Hemmady2012}. We use the same set of Hamiltonians and vary only the loss $\eta$ between ensembles, resulting in Fig.~\ref{Fig4}(b). 

RMT numerical results shown in panel (b) reveal a singularity density that has a near power law decrease with increasing uniform attenuation $\eta$. There are more intrinsic differences between our experimental ensembles than just loss, which is why the trend isn't as clear in panel (a). Overall, it seems that singularity abundance decreases with increasing loss. It should be noted that we are bounded above and below for values of $\eta$ we can simulate well with reasonable computation time. For $\eta>40$ we find very few (single digit number) singularities over an entire ensemble, and for $\eta<5$, the quality factor $Q$ of the resonances become so large that we need a much finer frequency spacing to ensure we properly detect any singularities. While the frequency resolution in RMT simulation is arbitrary, the computation time required is not.

\begin{figure*}[thb]
\includegraphics[width=\textwidth]{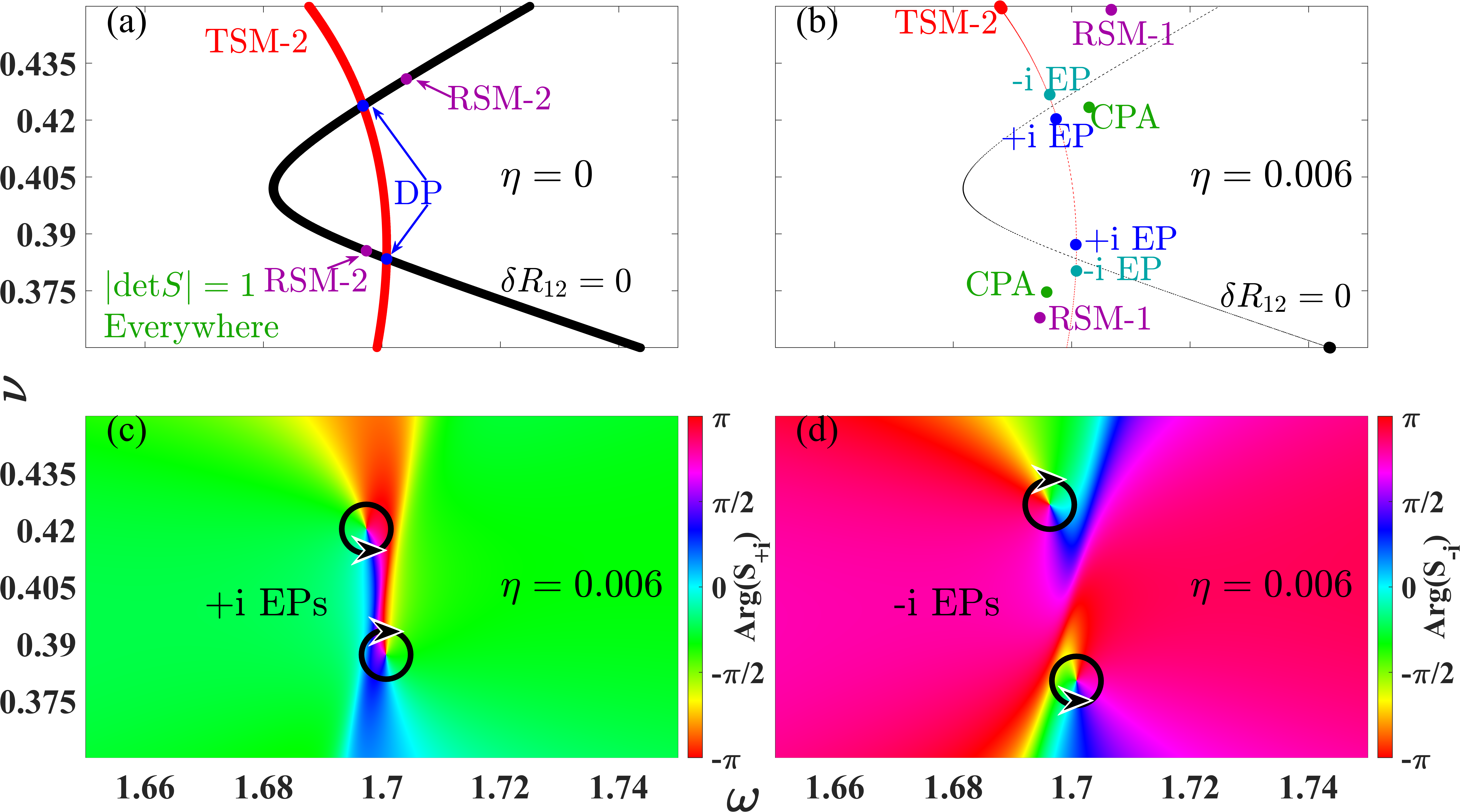}
\caption{(a) Locations of scattering zeros in the ($\omega,\nu$) parameter space of an RMT simulation with $\eta=0$. (b) Locations of scattering zeros in the ($\omega,\nu$) parameter space of an RMT simulation using the same effective Hamiltonian and mode-channel coupling as in (a) but with $\eta=0.006$. Dashed red and black lines show the $\eta=0$ locations of TSM-2 and $\delta R_{12}=0$ which have become \rev{single (orange and black)} points with the introduction of finite absorption. (c-d) Phase of $S_{+i}$ and $S_{-i}$ from simulation in (b). Black circles surround EPs (phase singularities) and arrows show direction of phase winding.}
\label{Fig5}
\end{figure*}

Note that at zero loss (i.e. a Hermitian system) certain singularities are forbidden, namely CPAs and EPs. Scattering parameter zeros (RSMs and TSMs) still occur because the elements of the $S$-matrix remain subunitary, as do symmetric reflection points and reciprocal points in non-reciprocal systems. However, the singularities can have different characteristics in a Hermitian system. Because we cannot experimentally measure a lossless system, we turn once more to RMT simulation. We consider the case of a reciprocal system with two channels ($\beta=1,M=2$) for simplicity using the model described in Ref.~\cite{erb2025} Appendix H. This model allows for Hamiltonian $H_0$ with a tunable parameter $\nu$ with the equation
\begin{equation}
H_0(\nu) = H_1 + \lvert cos(\nu) \rvert H_2 + \lvert sin(\nu) \rvert H_3
\label{Eqn_Hamiltonian}
\end{equation}
where $H_1,H_2$, and $H_3$ are all standard GOE random matrices. A scattering matrix that depends on both the parameter $\nu$ and a frequency $\omega$ can be calculated using coupled mode theory \cite{Fan2003}:
\begin{equation}
    S(\omega,\nu) = -I_{M\times M}+iWG(\omega,\nu)W^\dagger \nonumber
\end{equation}
where $G(\omega,\nu) = \frac{1}{\omega-H_0+\frac{i}{2}W^\dagger W}$ and the matrix $W$ describes the coupling between the $N$ modes of the closed system Hamiltonian $H_0$ to the $M$ scattering channels. Uniform absorption $\eta$ can be added to the model by making the simple substition $\omega\rightarrow\omega+\frac{i\eta}{4\pi}$.

% \begin{figure}[thb]
% \includegraphics[width=0.5\textwidth]{Fig5_v3.png}
% \caption{(a) Locations of scattering zeros in the parameter space of an RMT simulation with $\eta=0$. (b) Locations of scattering zeros in the parameter space of an RMT simulation using the same effective Hamiltonian and mode-channel coupling as in (a) but with $\eta=0.006$. Dashed red and black lines show the $\eta=0$ locations of TSM-2 and $\delta R_{12}=0$. (c-d) Phase of $S_{+i}$ and $S_{-i}$ from simulation in (b). Black circles surround EPs (phase singularities) and arrows show direction of phase winding.}
% \label{Fig5}
% \end{figure}

For a unitary $2\times2$ scattering matrix, the reflection parameters $S_{11}$ and $S_{22}$ can differ by up to a phase. Recalling that $\delta R_{xy}=S_{11}-S_{22}$, if $\text{Re}[\delta R_{12}]=0$, it must necessarily be the case that also $\text{Im}[\delta R_{12}]=0$. This makes the symmetric reflection condition $\delta R_{12}=0$ have the dimension of a curve as shown by the black line in Fig.~\ref{Fig5}(a). Further it means that if $|S_{11}|=0$, $|S_{22}|$ must also be zero so the reflection zeros are coincident and there are only RSM-2s, marked by the pink points in Fig.~\ref{Fig5}(a). It also happens to be the case that the TSM condition $S_{12}=0+i0$ has the dimension of a curve, which is the red line labeled TSM-2 in Fig.~\ref{Fig5}(a). The intersection of the $\delta R_{12}=0$ and TSM-2 curves results in a DP, as those locations are where the scattering matrix is exactly the identity matrix. The DPs are marked in dark blue.

\rev{We notice that in non-absorbing systems, what determines the abundance of singularities is more complicated than a single parameter ($\eta$) having the same effect on every singularity across different systems. Some singularities (e.g. CPAs and EPs) will not appear at all, while other singularities (e.g. symmetric reflection) may have certain properties (such as being a point or a curve in $d=2$ dimensional parameter space) depend on the number of channels $M$. The geometrical properties of the cavity, such as size and shape, can also determine if certain singularities are possible \cite{Mostafazedah2011,Mostafazadeh2012,Joshi2024}. Because of the high degree of variability between the properties of different singularities at zero loss (or equivalently, balanced absorption and gain), no universal statements can be made about singularity statistics in unitary scattering systems.}

In Fig.~\ref{Fig5}(b) we show the resulting singularities when we recalculate the $S$-matrix using the same model and parameters as in (a), but now with a small uniform attenuation $\eta=0.006$. We find that when a DP is perturbed in a way that introduces non-Hermiticity, it transforms into two EP-2s, \rev{similar to what was seen by Chen \textit{et al.} in Ref.~\cite{Chen2025}}. This is true for infinitesimally small amounts of loss, the smallest attempted in the simulations was $\eta=6\times 10^{-9}$, which can be seen in Appendix \ref{App_lowloss}. The charge of an EP-2 in a $M=2$ system is determined by whether the EP is a complex zero of $S_{+i} := S_{11} - S_{22} - 2i\sqrt{S_{12}S_{21}}$ (+i EPs marked in dark blue) or $S_{-i} := S_{11} - S_{22} + 2i\sqrt{S_{12}S_{21}}$ (-i EPs marked in light blue) \cite{erb2025}, the phase of which are shown in panels (c) and (d), respectively. The black circles and winding arrows show that each DP splits into a pair of EP-2s with opposite winding and charge \rev{$(\pm i$)}, such that the overall value of both these quantities is conserved when uniform loss is introduced.

In contrast, CPAs require some finite amount of loss to bring an $S$-matrix eigenvalue from the complex unit circle to zero. Exactly how much loss is required is a nuanced question since the trajectory of scattering eigenvalues in the complex plane is not linear. A lower bound, if it exists, should in principle be calculable, but is unknown to us. In the region of parameter space for the Hamiltonian used in Fig.~\ref{Fig5}, at $\eta=0.0006$ no CPAs could be found, however at $\eta=0.006$ a pair with opposite winding had already been created, marked by the green points in panel (b). With the addition of loss, not only have two CPAs appeared and the two DPs split into four EPs, but also the RSM-2s separated into RSM-1s (the pair corresponding to $|S_{22}|=0$ have already annihilated by $\eta=0.006$ so only two remain), and the $\delta R_{12}$ and TSM-2 lines have broken up into \rev{black and orange single} points, \rev{respectively}.

Figs.~\ref{Fig4}-\ref{Fig5} raise several questions that are outside the scope of this paper. Since the number of CPAs in a lossless system is zero, there must be some critical amount of loss $\eta_c$ at which point the number of CPAs is maximized before it starts to decrease, as we see in Fig.~\ref{Fig4}. Is there a way to estimate $\eta_c$? In contrast, EP-2s exist in abundance at infinitesimal loss as each DP splits into two EP-2s with the addition of absorption, as shown in Fig.~\ref{Fig5}.  But for $0<\eta\leq1$ we don't know what happens to the number of Exceptional Points. Does it increase, remain stable, or monotonically decrease? Scattering parameter zeros are similarly always present and the same questions can be asked about them. The fact that the four symbols for the different singularities in Fig.~\ref{Fig4} maintain such close proximity to each other, both in the experimental data and RMT numerics, suggests that all singularities are approximately equally plentiful. Does this extend to lower loss? If there is a critical amount $\eta_c$ which maximizes the density of CPAs, does this same amount of loss maximize the density of all other singularities as well? 

\subsection{Higher Order Singularities - Statistics} \label{Sub_HighStats}
\begin{figure}[thb]
\includegraphics[width=0.48\textwidth]{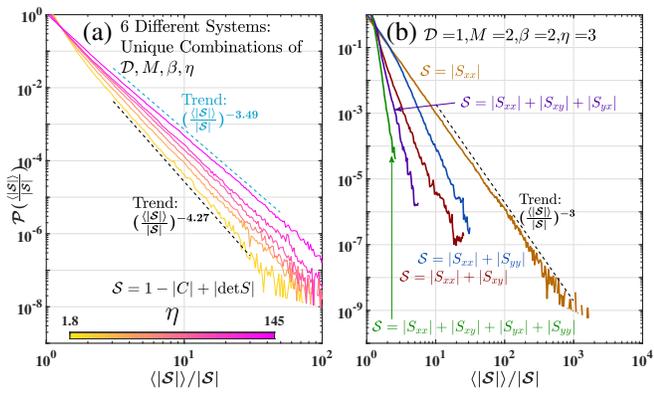}
\caption{(a) PDFs of $\frac{\langle |\mathcal{S}|\rangle}{|\mathcal{S}|}$ for $\mathcal{S} = 1-|C|+|\text{det}S|$ from 6 ensembles of experimental scattering systems with different values for the four parameters $\mathcal{D},M,\beta,\eta$.  Color gradient of curves corresponds to different values of $\eta$ (not uniformaly spaced) of the ensembles. (b) PDFs of $\frac{\langle |\mathcal{S}|\rangle}{|\mathcal{S}|}$ for various functions $\mathcal{S}$ from an arbitrary ensembles.}
\label{Fig6}
\end{figure}

Finally, we consider the question of higher order singularities, which are composites of multiple fundamental singularities. Through the use of tunable perturbers, it is possible to make independent singularities coincident in parameter space, which results in novel applications such as the 50:50 IQ power splitter described in \cite{erb2025} using a coincident CPA-EP-2.  The intersection of complex zeros of different fields has no topological stability, and the singularities can move apart under arbitrarily small perturbation. Consequently, we do not find a universal power law for all PDFs of functions that diverges at a singularity of order $P>2$, for any given $P$.

Fig.~\ref{Fig6}(a) demonstrates the experimental PDFs of $\frac{\langle |\mathcal{S}|\rangle}{|\mathcal{S}|}$ for $\mathcal{S} = 1-|C|+|\text{det}S|$ which diverge at a combined CPA-EP-2 of the $S$-matrix. The different curves come from 6 ensembles with different values for the parameters $\mathcal{D},M,\beta,\eta$. The bright pink curve, which comes from the lossiest system measured ($\eta=145$), has a power law fit of $-3.49$ as demonstrated by the dashed blue line, while the bright yellow curve which comes from the system with the least absorption ($\eta=1.8$) has a power law fit of $-4.27$, demonstrated by the dashed black line. The four other curves at intermediate values of uniform attenuation fill out the space in between with slopes bounded by those two values. This shows that even though $\mathcal{P}(\frac{\langle|\text{det}S|\rangle}{|\text{det}S|})$ and $\mathcal{P}(\frac{\langle1-|C|\rangle}{1-|C|})$ have universal power laws, the power law of a PDF of a higher order combined singularity is system detail-dependent. It is interesting to note that the steepest power law comes from the ensemble with the lowest loss and increasing absorption makes the PDF drop slower. This implies that although the density of individual singularities decreases with increasing loss, their intersections become more common. It is currently unclear why this would be the case.

Panel (b) of Fig.~\ref{Fig6} has PDFs of five different quantities from an experimental non-reciprocal ($\beta=2$) graph ($\mathcal{D}=1$) with $M=2$ channels and $\eta=3$. In orange is $\mathcal{P}(\frac{\langle |S_{xx}|\rangle}{|S_{xx}|})$ for reference. The blue curve is the PDF that combines $|S_{xx}|$ and $|S_{yy}|$ which shows the abundance of RSM-2s. Similarly, the red curve combines $|S_{xx}|$ and $|S_{xy}|$ which shows the abundance of $S$-matrix row zeros. Despite the overall non-reciprocity of the system, statistically row and column zeros behave the same so we do not show a PDF for $|S_{xx}|$ and $|S_{yx}|$. The purple curve combines $|S_{xx}|$, $|S_{xy}|$ and $|S_{yx}|$ to show the probability of a simultaneous row and column zero. This curve is steeper than either the red or the blue ones, as the addition of another constraint dramatically reduces the likelihood of such an event. A third order singularity like this is so improbable that the PDF from this ensemble does not get past $\frac{\langle |\mathcal{S}|\rangle}{|\mathcal{S}|}=6$ which means that definitely none of these singularities were captured in this ensemble. The green curve examines the case of a zero of the full $2\times2$ scattering matrix, and is therefore even steeper, and ends earlier than the purple curve. The more constraints, the higher the order of the singularity being considered, the harder it is to get to the tail of the PDF to identify a power law.

A further statement should be made at this stage about higher order singularities of the scattering matrix. The fundamental singularities being combined are of not wholly independent complex scalars. CPAs are zeros of $\text{det}S$, but EPs also depend on the principal invariants of the $S$-matrix, one of which is its determinant. Similarly, $S_{xx}$ and $S_{xy}$ have  inherent correlations which can depend on many factors, including the number of channels $M$ or loss $\eta$, as well as more complicated issues not considered in this paper such as imperfect mode-channel coupling. It is possible that truly universal statistics exist for composites of fully independent singularities, but \rev{it is very difficult to construct multiple complex scalar functions with no correlations out of a physical scattering matrix.}

\section{Conclusion}\label{Sec_Conc}
We have examined the statistical properties of $S$-matrix singularities by generalizing the concept of singularity speckle patterns to arbitrary two dimensional parameter spaces and any complex scalar function that describes wave phenomena involving complicated scattering. The statistics discussed come from experimental microwave resonant systems with the number of channels ranging from $M=1$ to $M=4$. Both systems with and without Lorentz reciprocity were considered, as well as a wide range of uniform absorption ($1.8\leq\eta\leq145$). Supporting RMT simulations were conducted with results that agree with the empirical data, showing these findings are generic to all wave scattering systems, such as acoustic, optical, and photonic resonators. Fundamental singularities, which are topologically-protected vortices of complex scalar fields in two-dimensional parameter space, have a ``superuniversal'' statistical rule that a function which \rev{has a simple pole} at a singularity has a PDF with a -3 power law tail.  These PDFs can then be used to estimate singularity abundance based on system parameters, and a general rule \rev{for absorbing systems is} an increase in absorption leads to a decrease in singularity density. Higher order singularities, which are the degeneracies of fundamental singularities, are not topologically stable and we do not see any universal statistical properties for them. 

Several questions about how the singularity abundance depends on absorption have been raised by these results and are suggested as future work. The statistical ``superuniversality'' seen here may depend on having perfect coupling as was found for complex time delay PDFs in Ref.~\cite{shaibe2025}. How coupling between a scattering environment and the channels leading in and out affects the universal power laws \rev{and} singularity density presented in this paper, are other unexplored problems. There are also opportunities for similar analysis in other fields of physics, as complex scalar fields or order parameters are commonly studied quantities.

%\section{Acknowledgements}
\bigskip
\textbf{Acknowledgements} We acknowledge helpful discussions with Thomas M. Antonsen. We thank David Shrekenhamer and Timothy Sleasman of JHU APL for the design and fabrication of the metasurfaces used in this work. This work was supported by NSF/RINGS under grant No. ECCS-2148318, ONR under grant N000142312507, and DARPA WARDEN under grant HR00112120021.  

\appendix

\section{Simple Mathematical Reason for Universal Power Laws} \label{App_Proof}

Every fundamental, first order singularity can be written as a zero of some complex scalar function $\mathcal{S}=u+iv=re^{i\theta}$. If the PDF  $\mathcal{P}_\theta(\theta)$ is uniform, a reasonable assumption to make and true for all functions $\mathcal{S}$ considered in this paper, then the probability density function of $\mathcal{S}$ in polar coordinates can be written
\begin{equation}
\mathcal{P}^{\text{polar}}_\mathcal{S}(r,\theta)=\frac{1}{2\pi}\mathcal{P}_r(r). \label{EqS1}
\end{equation}

We desire the PDF of $\mathcal{T}=\rho e^{i\phi}$ where $\mathcal{T}=\frac{1}{\mathcal{S}}$. If there is symmetry in $\theta$ then there is symmetry in $\phi=-\theta$, hence we can write 
\begin{equation}
   \mathcal{P}^{\text{polar}}_\mathcal{T}(\rho,\phi)= \frac{1}{2\pi}\mathcal{P}_{\rho}(\rho). \label{EqT1}
\end{equation}
We perform a standard change of variable trick using a Dirac delta function to compare the right hand sides of Eqs.~\ref{EqS1}-\ref{EqT1}:
\begin{equation}
    \mathcal{P}_{\rho}(\rho)=\int\mathcal{P}_r(r)*\delta\left(\rho-\frac{1}{r}\right)*rdr. \nonumber
\end{equation}
The delta function can be transformed in the following manner
\begin{equation}
    \delta\left(\rho-\frac{1}{r}\right)=\delta\left(r-\frac{1}{\rho}\right)*\left|\frac{d\rho}{dr}\right|^{-1}=\delta\left(r-\frac{1}{\rho}\right)\frac{1}{\rho^2} \nonumber
\end{equation}
which when plugged into the above expression returns
\begin{equation}
\begin{split}
        \mathcal{P}_{\rho}(\rho)&=\int\mathcal{P}_r(r)*\delta\left(r-\frac{1}{\rho}\right)\frac{1}{\rho^2}*rdr  \\
        &=\frac{1}{\rho^3}\mathcal{P}_r(\frac{1}{\rho}) \label{EqPow1}
\end{split}
\end{equation}
which has the empirically found $-3$ power law. Something similar can be done to show that the PDF $\mathcal{P}_w(w)$ has a $-2$ power law, where $w=\frac{1}{u}$. The steps are the same:
\begin{equation}
    \mathcal{P}_w(w)=\int\mathcal{P}_u(u)*\delta\left(w-\frac{1}{u}\right)*du \nonumber
\end{equation}
\begin{equation}
    \delta\left(w-\frac{1}{u}\right)=\delta\left(u-\frac{1}{w}\right)*\left|\frac{dw}{du}\right|^{-1}=\delta\left(u-\frac{1}{w}\right)\frac{1}{w^2} \nonumber
\end{equation}
\begin{equation}
\begin{split}
        \mathcal{P}_{w}(w)&=\int\mathcal{P}_u(u)*\delta\left(u-\frac{1}{w}\right)\frac{1}{w^2}*du  \\
        &=\frac{1}{w^2}\mathcal{P}_u(\frac{1}{w}) \label{EqPow2}
\end{split}
\end{equation}
and results in the expected $-2$ power law.

\begin{figure}[bht]
\includegraphics[width=0.48\textwidth]{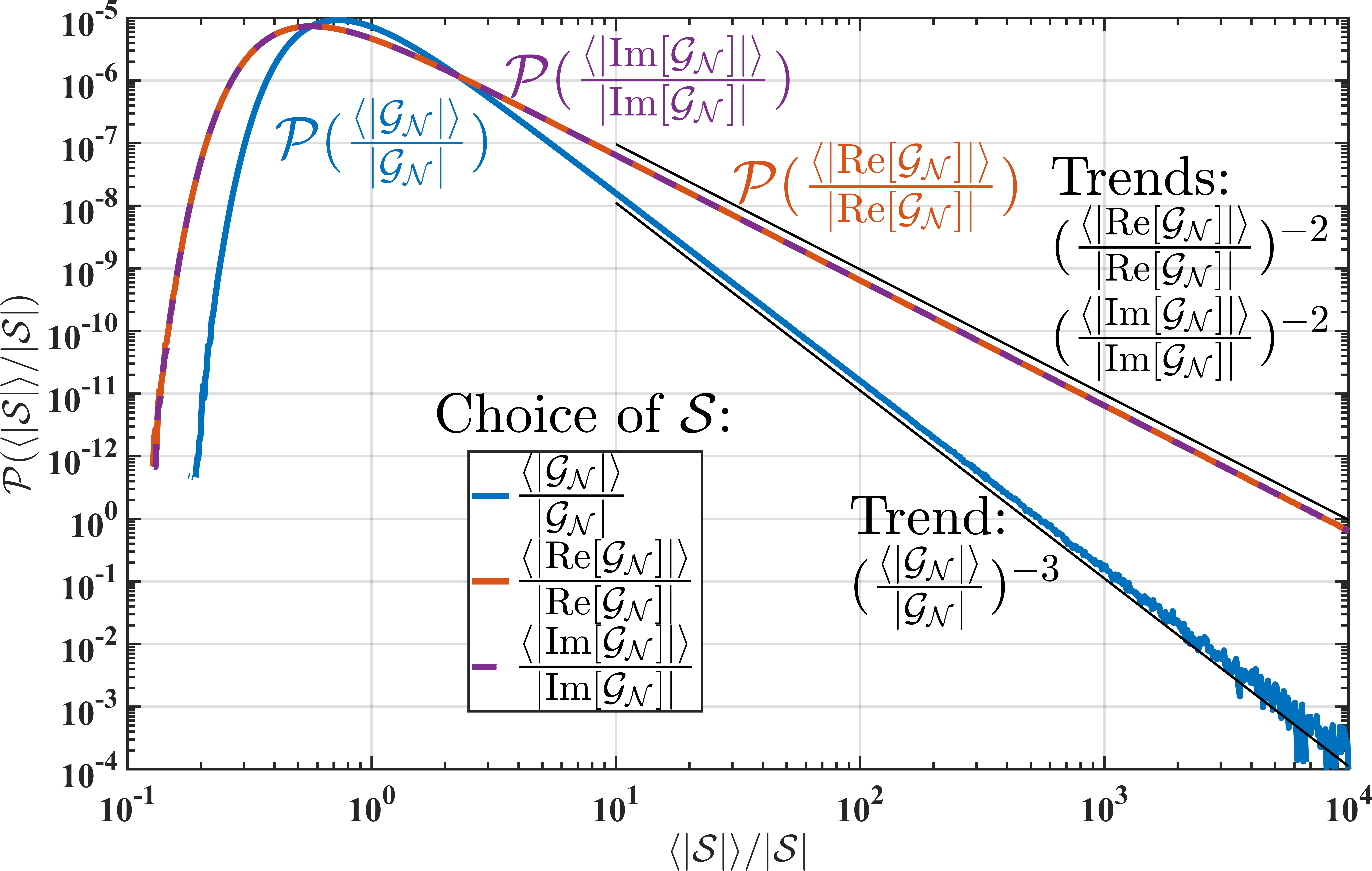}
\caption{PDFs of the inverse of complex Gaussian random numbers $\mathcal{G_N}$. These PDFs show the same power law tails as the experimental data and RMT numerics: $-3$ for $\mathcal{P}(\frac{\langle|\mathcal{G_N}|\rangle}{|\mathcal{G_N}|})$ and $-2$ for $\mathcal{P}(\frac{\langle|\mathrm{Re}[\mathcal{G_N}]|\rangle}{|\mathrm{Re}[\mathcal{G_N}]|})$ and $\mathcal{P}(\frac{\langle|\mathrm{Im}[\mathcal{G_N}]|\rangle}{|\mathrm{Im}[\mathcal{G_N}]|})$.}
\label{App_Rand}
\end{figure}

To show this is purely a mathematical result which requires no physics, we generated $10^9$ complex numbers $\mathcal{G_N}$ from a Gaussian normal distribution. Fig.~\ref{App_Rand} shows the PDFs of $\frac{\langle|\mathcal{G_N}|\rangle}{|\mathcal{G_N}|}$, $\frac{\langle|\mathrm{Re}[\mathcal{G_N}]|\rangle}{|\mathrm{Re}[\mathcal{G_N}]|}$, and $\frac{\langle|\mathrm{Im}[\mathcal{G_N}]|\rangle}{|\mathrm{Im}[\mathcal{G_N}]|}$. Indeed, we see the same $-3$ and $-2$ power laws as in our experimental scattering matrix data.

\section{\rev{Other Power Law Tails}} \label{App_Tails}

\rev{As discussed in the paper, the condition for the PDF of a quantity to have a $-3$ ($-2$) power law tail is that the quantity diverges as a simple pole at a singularity (one degree of freedom constraint of a complex scalar function). If the divergence does not have the form of a simple pole, the tail of the PDF will not have a $-3$ power law, in general.}

\begin{figure}[bht]
\includegraphics[width=0.48\textwidth]{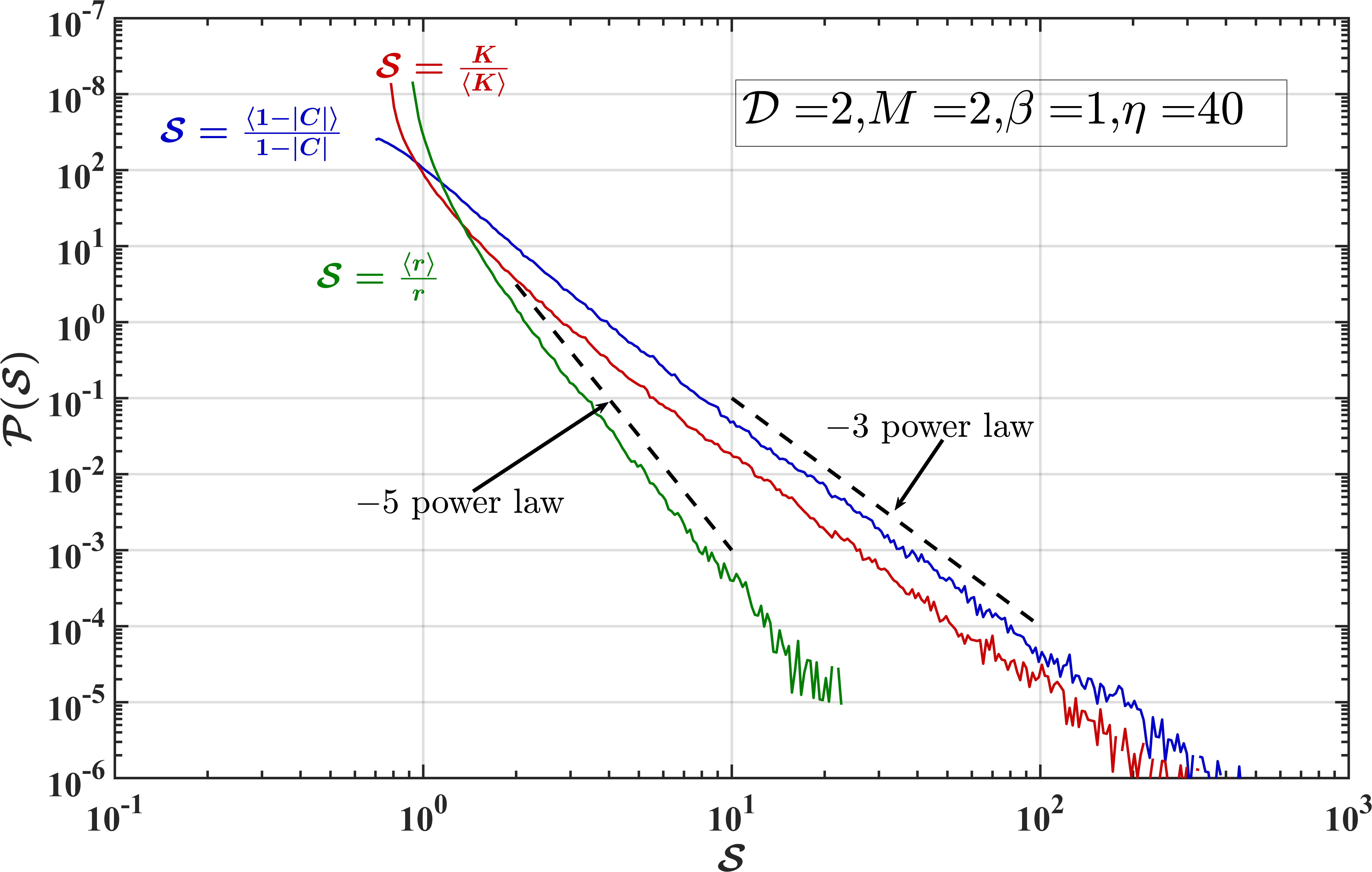}
\caption{\rev{PDFs of one over one minus the eigenvector coalescence $(1-|C|)^{-1}$, Petermann factor $K$, and inverse condition number $r^{-1}$ from the same experimental ensemble as panel (c) of Fig.~\ref{Fig3} in the main text. All three of these quantities diverge solely at EP-2s. Dashed black lines characterize the power laws of the tail behaviors.}}
\label{App_CKr}
\end{figure}

\begin{figure}[bht]
\includegraphics[width=0.48\textwidth]{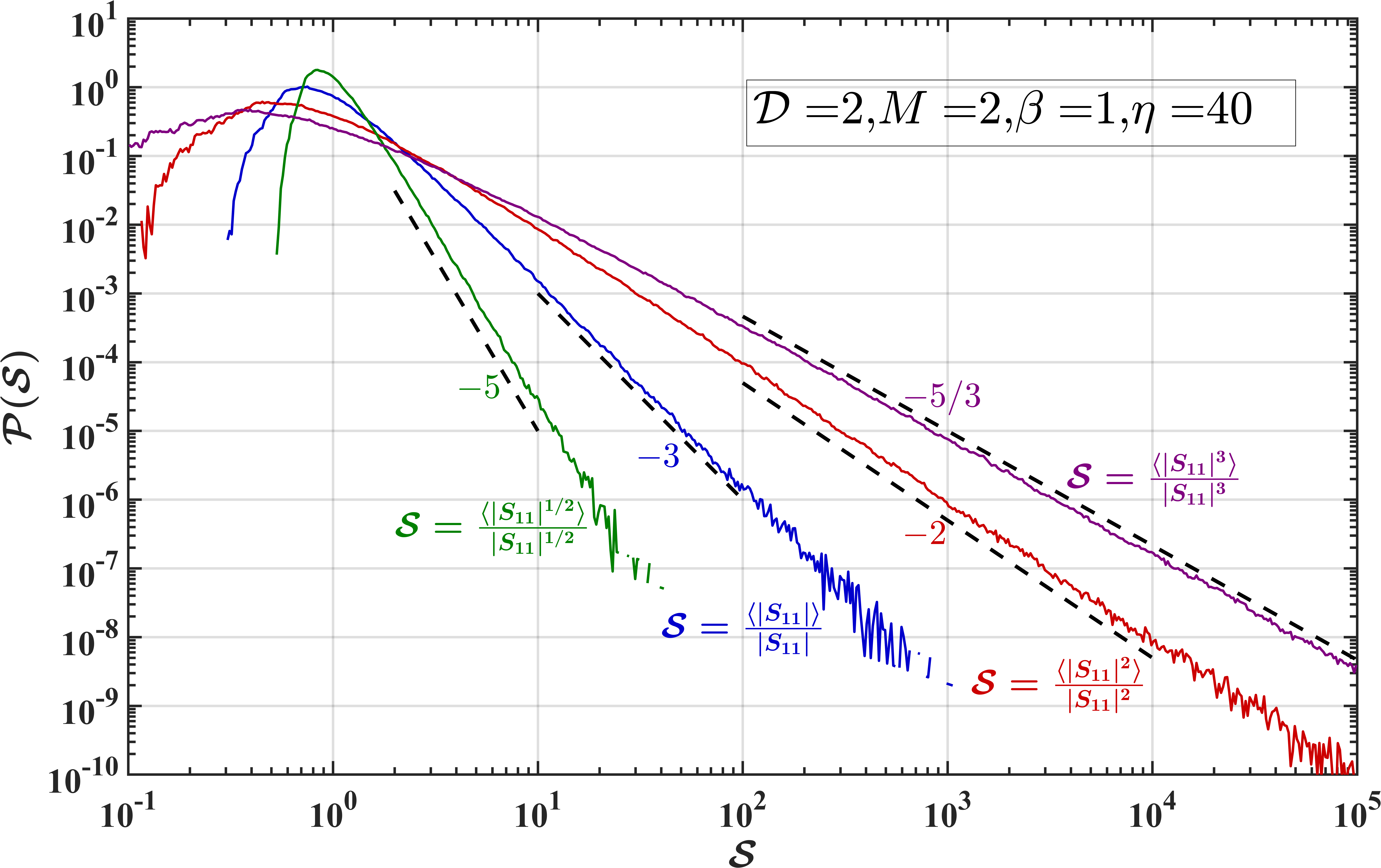}
\caption{\rev{PDFs of $\frac{\langle |S_{11}| \rangle}{|S_{11}|}$ raised to different powers. The $S_{11}$ data comes from the same experimental ensemble as panel (c) of Fig.~\ref{Fig3} in the main text. All the quantities diverge solely at $S_{11}=0+i0$ (RSM-1). Dashed black lines characterize the power laws of the tail behaviors.}}
\label{App_PWRTAILs}
\end{figure}

\rev{As mentioned before, the condition number $r$ has a branch point at an EP-2  since it can be written as $r=\sqrt{1-|C|^2}$ \cite{Guo2023}. Therefore, even though $\frac{1}{r}$ diverges at an EP-2, the same as $\frac{1}{1-|C|}$ and the Petermann factor $K$, the PDF $\mathcal{P}(\frac{\langle r \rangle}{r})$ has a tail with a $-5$ power law, not $-3$. In Fig.~\ref{App_CKr}, we show the experimental PDFs for these three quantities, where the data comes from the same $\mathcal{D}=2,M=2,\beta=1,\eta=40$ ensemble as panel (c) of Fig.~\ref{Fig3}. To further illustrate that the $-5$ power law results from the way $\frac{1}{r}$ diverges at an EP-2, we show the PDFs of $\frac{\langle |S_{11}| \rangle}{|S_{11}|}$ raised to different powers in Fig.~\ref{App_PWRTAILs}. We label each PDF by the power law of the tail and find that $\mathcal{P}(\frac{\langle |S_{11}|^{1/2} \rangle}{ |S_{11}|^{1/2}})$ has the same $-5$ power law as $\mathcal{P}(\frac{\langle r \rangle}{r})$. Notably, $\frac{1}{ |S_{11}|^{1/2}}$ has a branch point at an RSM-1, just like $\frac{1}{r}$ at an EP-2. The red and purple curves ($\mathcal{P}(\frac{\langle |S_{11}|^{2} \rangle}{ |S_{11}|^{2}})$ and $\mathcal{P}(\frac{\langle |S_{11}|^{3} \rangle}{ |S_{11}|^{3}})$, respectively) further show that the PDFs of functions that have higher order poles at singularities also do not have $-3$ power law tails.}

\section{Singularities in $d=3$ Parameter Space} \label{App_3ParaSpace}

As stated in Sec.~\ref{Sec_Intro}, while singularities form speckle patterns of points in $d=2$ dimensional parameter space, in $d=3$ they form continuous curves because singularities always have dimension $d-2$. These curves either form closed loops or trail off the edge of the explored parameter space, with no end points visible. Note that this is the same property we see for the $d-1$ dimensional curves in a $d=2$ dimensional parameter space. In $d=3$ space, the $d-1$ dimensional topologically protected structures such as $\text{Re}[\text{det}S]=0$ take the form of surfaces that are either closed or extend beyond the limits of the explored parameter space. In simulations, it can be seen that sometimes $d-1$ structures in $d=2$ space extend out to infinite value of certain parameters (such as loss $\eta$), so we assume the same must be true for $d-2$ structures in higher dimensional spaces. Experimentally, our tunable perturbations are limited within some range, so we cannot verify this assumption.

\begin{figure}[bht]
\includegraphics[width=0.48\textwidth]{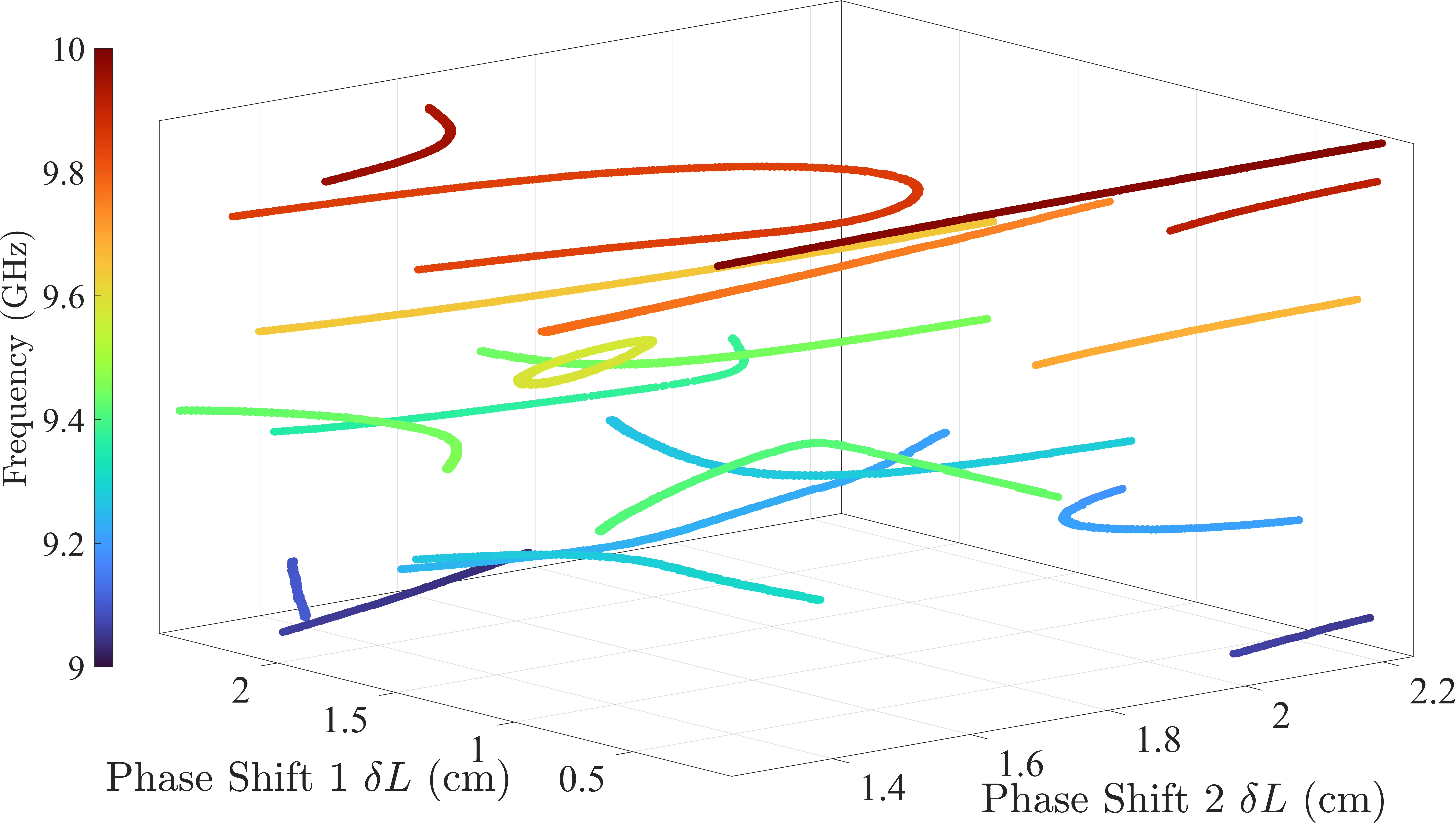}
\caption{ Three-dimensional parameter space view of singularity lines for $\text{det}S=0$ (CPA) from an experimental reciprocal, two channel microwave graph ($\beta=1,M=2,\mathcal{D}=1$) measurement. Different colors correspond to different frequency values. The curves do not start or stop anywhere within the interior of the measured parameter space, only on the edges.}
\label{App_3Para}
\end{figure}

In Fig.~\ref{App_3Para}, we show the locations in $d=3$ dimensional parameter space of CPA singularities ($\text{det}S=0$) from an experimental reciprocal two channel graph ($\beta=1,M=2,\mathcal{D}=1$). Because this is not an ensemble measurement, and the global properties of the graph are different at every point (two phase shifters changing resulting in a varying total length of the graph), there is no single $\eta$ value that can be assigned to this data. At best we can say that $2\leq\eta\leq3$ everywhere in this parameter space based on ensemble measurements using similar lengths in this frequency range. The olive green curve which corresponds to $\sim9.58$ GHz forms a closed loop, while every other curve extends to the edges of the included parameter space.

In Fig.~\ref{App_3Para}, there is a color gradient in frequency but visually every line appears to be a single solid color. Most scattering singularities, CPAs included, do not move much in frequency under perturbations to a graph system in the form of bond length changes. In $\mathcal{D}=2$ and $3$ dimensional cavities there are more pronounced movements in frequency, though still nothing dramatic. Practically, this means it is easier to move multiple different S-matrix zeros to different spots along the real frequency axis than to drag a single zero to different spots along the real frequency axis. For graphs, most frequency movement is on the order of the mean mode spacing or less, which for this graph is about $40$ MHz. An example of this can be seen in Fig.~4(a) of Ref.~\cite{erb2025robustwavesplittersbased}, which shows a CPA singularity line extending over $60$ MHz, the data coming from the same graph used in Fig.~\ref{App_3Para}.

\section{\rev{Experimental Ensemble Details}} \label{App_ENS}

\rev{The statistical results presented in this paper were obtained from 31 ensembles of experimental microwave systems, each with a unique combination of the four parameters $\mathcal{D},M,\beta,\eta$. In Table~\ref{ENS_Table}, we provide the parameter values for these ensembles, as well as the estimated average number of modes in each realization and the total number of realizations per ensemble. Note that each realization is made up of either $64000$, $80000$, or $100000$ (depending on the capabilities of the network analyzer used) equally spaced frequency points so that we can well characterize the modes and ensure we identify extreme events such as scattering singularities. Each ensemble is therefore over $2.5$ GB of independent scattering matrix data. Details of how the systems are perturbed to create ensembles are discussed in Sec.~\ref{Sec_Exp}.}

\begin{table}
    \centering
    \begin{tabular}{|P{0.6cm}|P{0.6cm}|P{0.6cm}|P{3cm}|P{1.5cm}|}
    \hline
        \rev{$\mathcal{D}$} & \rev{$M$} & \rev{$\beta$} & \rev{$\eta$ Values (Estimated Number of Modes)} & \rev{Number of Realizations}\\
        \hline
        1 & 1 & 1 & 2.4 (116), 7.3 (121) & 715 for all ensembles \\
        \hline
        1 & 1 & 2 & 3.1 (119), 7.5 (124) & 715 for all ensembles \\
        \hline
        1 & 2 & 1 & 1.8 (43), 2.1 (65), 2.4 (116), %
        2.5 (43), 2.8 (65), 4.4 (84), 5.7 (84), 7.0 (84),%
        7.3 (122), 10.1 (84), 11.3 (86), 16.3 (86), %
        21.4 (86), 30.2 (86) &
        715 for all ensembles \\
        \hline
        1 & 2 & 2 & 1.9 (44), 2.9 (44), 3.1 (119), 7.5 (125) & 715 for all ensembles \\
        \hline
        1 & 3 & 1 & 2.5 (44) & 715 for all ensembles \\
        \hline
        1 & 3 & 2 & 3.1 (45), 3.8 (47) & 715 for all ensembles\\
        \hline
        2 & 2 & 1 & 39.0 (68), 40.2 (145), 41.5 (76) & 302, 324, 301\\
        \hline
        2 & 3 & 1 & 40.2 (145)  & 196\\
        \hline
        3 & 2 & 1 & 49.0 (1000) & 804 \\
        \hline
        3 & 3 & 1 & 49.0 (520), 69.1 (1100), 79.2 (1400), 93.0 (1700), 145 (2100) & 804 for all ensembles \\
        \hline
    \end{tabular}
    \caption{\rev{Table of experimental ensemble parameters. The first three columns correspond to system dimension $\mathcal{D}$, number of channels $M$, and symmetry class $\beta$. Since there are often multiple ensembles that have the same values of $\mathcal{D},M$, and $\beta$, these columns are shared for the sake of compactness. The fourth column has the ensemble value of the cavity absorption $\eta$ (as well as the estimated number of modes in the bandwidth measured). The number of modes is estimated by the bandwidth divided by the mean mode spacing $\Delta$. The final column has the number of realizations in the ensemble.}}
    \label{ENS_Table}
\end{table}

\section{Singularity Statistics - Random Matrix Theory} \label{App_RMT}

\begin{figure*}[hbt]
\includegraphics[width=\textwidth]{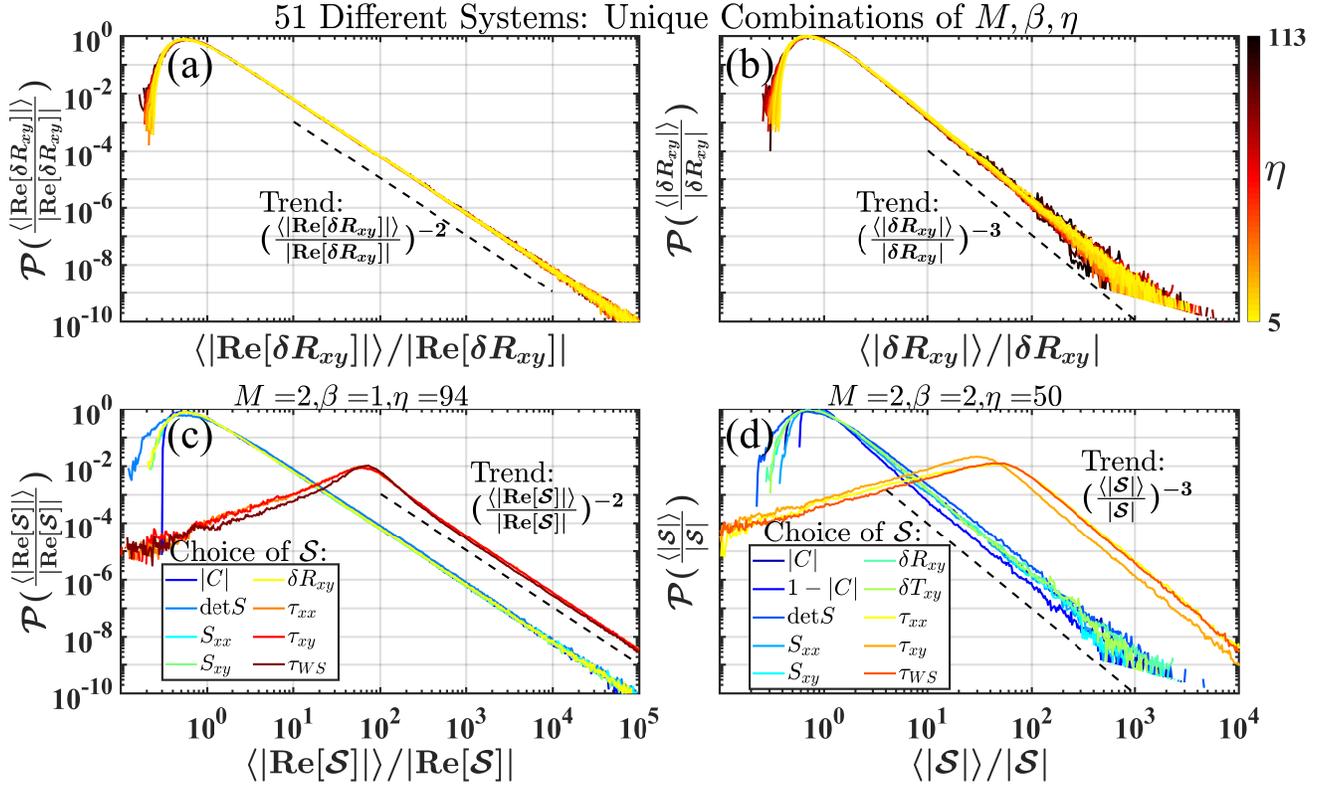}
\caption{(a-b) PDFs of $\frac{\langle |\text{Re}[\delta R_{xy}]| \rangle}{|\text{Re}[\delta R_{xy}]|}$ and $\frac{\langle |\delta R_{xy}| \rangle}{|\delta R_{xy}|}$ from 51 RMT ensembles with different values for the three parameters $M,\beta,\eta$.  Color gradient of curves corresponds to $\eta$ (not uniformly spaced) of the ensembles. Dashed black lines characterize \rev{the} power law of \rev{the} tail behavior. (c-d) PDFs of $\frac{\langle|\text{Re}[\mathcal{S}]| \rangle}{|\text{Re}[\mathcal{S}]|}$ and $\frac{\langle |\mathcal{S}|\rangle}{|\mathcal{S}|}$ for various functions $\mathcal{S}$ from two arbitrary RMT ensembles. Dashed black lines characterize \rev{the} power law of \rev{the} tail behavior.}
\label{App_RMT_PDF}
\end{figure*}

The process used to arrive at statistical results from RMT is as follows. First, we create two sets of $600$ independent Hamiltonians that are of dimension $10^5\cross10^5$, generated using GOE (for $\beta=1$) and Gaussian Unitary Ensemble (GUE) (for $\beta=2$) algorithms. Then normalized impedance matrices $Z$ are calculated using the Random Coupling Model \cite{Hemmady2005,Zheng20052Port,Zheng2006,shaibe2025} with the only parameter changed between different ensembles being the dimensionless uniform attenuation $\eta$. The impedance matrices are converted to scattering matrices through the formula:
\begin{equation}
    S=\frac{Z-Z_0I_{M\times M}}{Z+Z_0I_{M\times M}}. \nonumber
\end{equation}

 Note that because the impedance matrices created through RCM are normalized, the characteristic impedance $Z_0=1$. The result is a collection of ensembles of $600$ independent $S$-matrices, each ensemble characterized by $M,\beta$ and $\eta$. We treat these scattering matrices generated through RMT the same as our experimental ensembles and conduct the same analysis to calculate singularity mode density, shown in Fig.~\ref{Fig4}(b).

The RMT data can also be used to calculate PDFs of various complex scalar functions. In Fig.~\ref{App_RMT_PDF}, we show the same PDFs as in Fig.~\ref{Fig3}: (a) $\frac{\langle |\text{Re}[\delta R_{xy}]| \rangle}{|\text{Re}[\delta R_{xy}]|}$, (b) $\frac{\langle|\delta R_{xy}| \rangle}{|\delta R_{xy}|}$, and the same eight and ten functions in panels (c) and (d), respectively. The simulations utilize two sets of $600$ independent Hamiltonians that are of dimension $10^5\cross10^5$, generated using GOE (for $\beta=1$) and Gaussian Unitary Ensemble (GUE) (for $\beta=2$) algorithms. Then the scattering matrices are calculated using the Random Coupling Model \cite{Hemmady2005,Zheng20052Port,Zheng2006,shaibe2025} with the only parameter changed between different ensembles being the dimensionless uniform attenuation $\eta$.

Identical behavior is seen in Fig.~\ref{App_RMT_PDF} as in Fig.~\ref{Fig3}: $-2$ power law tails are seen panels (a) and (c), and $-3$ power law tails are seen in panels (b) and (d). This shows that even though the data in this paper is measured from microwave cavities, the results we present are true for all chaotic wave scattering systems. The present paper also explains why in the Appendix of Ref.~\cite{shaibe2025}, a $-2$ power law tail was seen for the probability distributions of complex reflection and transmission time delay differences ($\tau_{\delta R}$ and $\tau_{\delta T}$) from Hermitian ($\eta=0$) systems. As shown in Fig.~\ref{Fig5}(a), in a Hermitian system, $\delta R_{12}=0$ is $d-1$ dimensional, not a $d-2$ dimensional singularity. The same is true of $\delta T_{12}=0$ if the system is also non-reciprocal. Further the PDFs of complex transmission time delay from a Hermitian, reciprocal system would have $-2$ power laws, which is why the authors of Ref.~\cite{shaibe2025} chose to show plots from $\beta=2$ RMT data, since at the time they did not have the topology argument to understand why they did not see $-3$ power laws.

A final comment is that the PDF of $\frac{\langle|C|\rangle}{|C|}$, where $|C|$ is the coalescence of the $S$-matrix eigenvectors, from the reciprocal system shown in panel (c) has the expected $-2$ power law tail out to the largest values, never transitioning to $-3$ as seen in the experimental data in Fig. \ref{Fig3}(c). This is because RMT numerics lacks the small amount of residual non-reciprocity we always see in experimental data.

\section{Exceptional Points at Infinitesimal Absorption} \label{App_lowloss}

\begin{figure}[bht]
\includegraphics[width=0.5\textwidth]{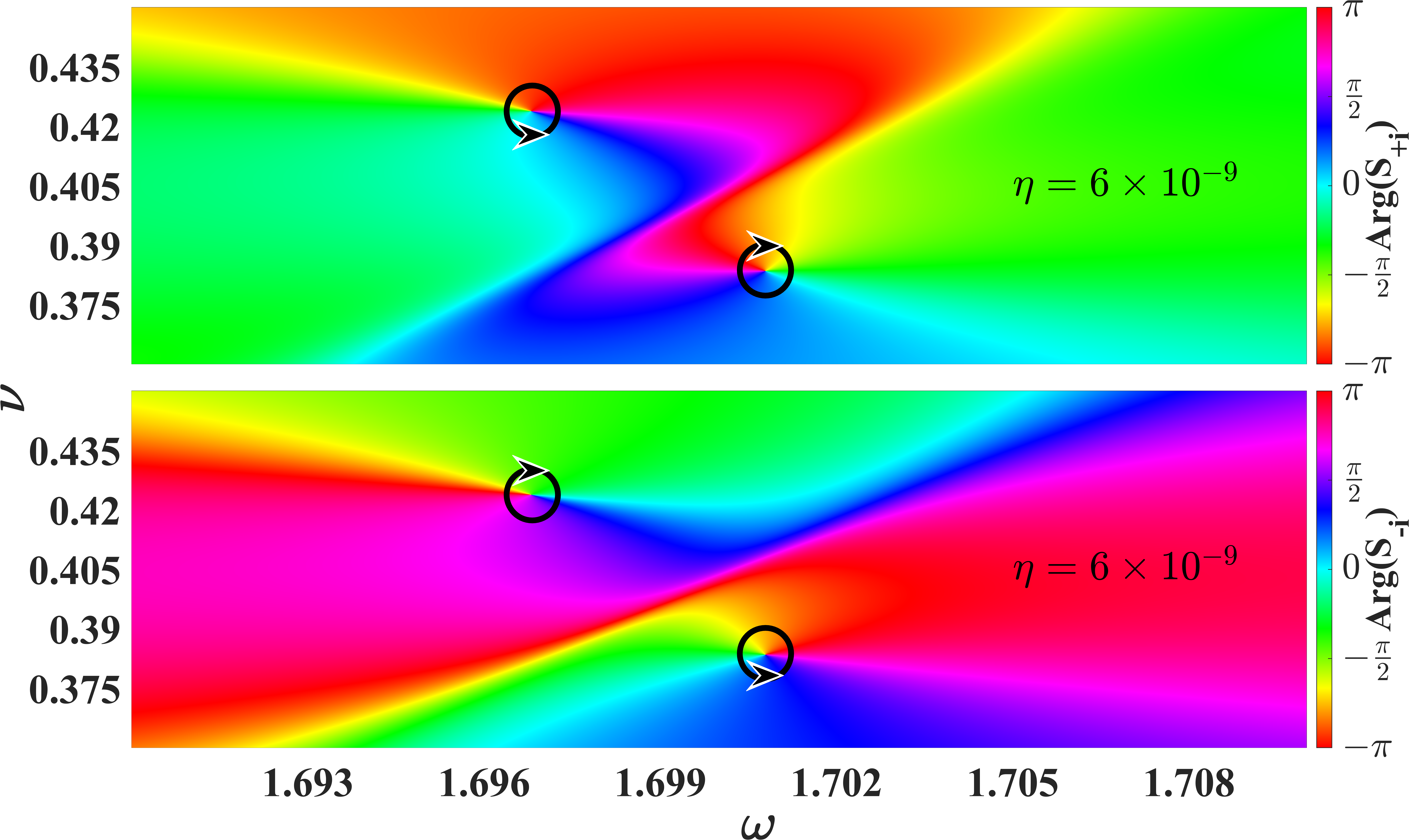}
\caption{Phase of $S_{+i}$ and $S_{-i}$ from the same kind of simulation as in Fig.~\ref{Fig5} over the same parameter space, but with smaller $\eta=6*10^{-9}$. Black circles surround EPs (phase singularities) and arrows show direction of phase winding.}
\label{App_EpsLoss}
\end{figure}

The addition of any amount of absorptive loss to a Hermitian scattering system causes DPs to split into two, oppositely charged EP-2s with opposite winding. Using the same Hamiltonian and parameter space as in Fig.~\ref{Fig5}, we reproduce panels (c) and (d) but with $\eta=6\times10^{-9}$, as shown in Fig.~\ref{App_EpsLoss}. 

In this way, Exceptional Points are unique among scattering singularities in that they exist at infinitesimal loss but not zero loss. All other scattering singularities we have found are either already present at zero loss, such as RSMs, or have a sufficient loss requirement before they are possible, such as CPAs.

\section{Complex Time Delay - Generalizations} \label{App_CTD}

\begin{figure*}[hbt]
\includegraphics[width=\textwidth]{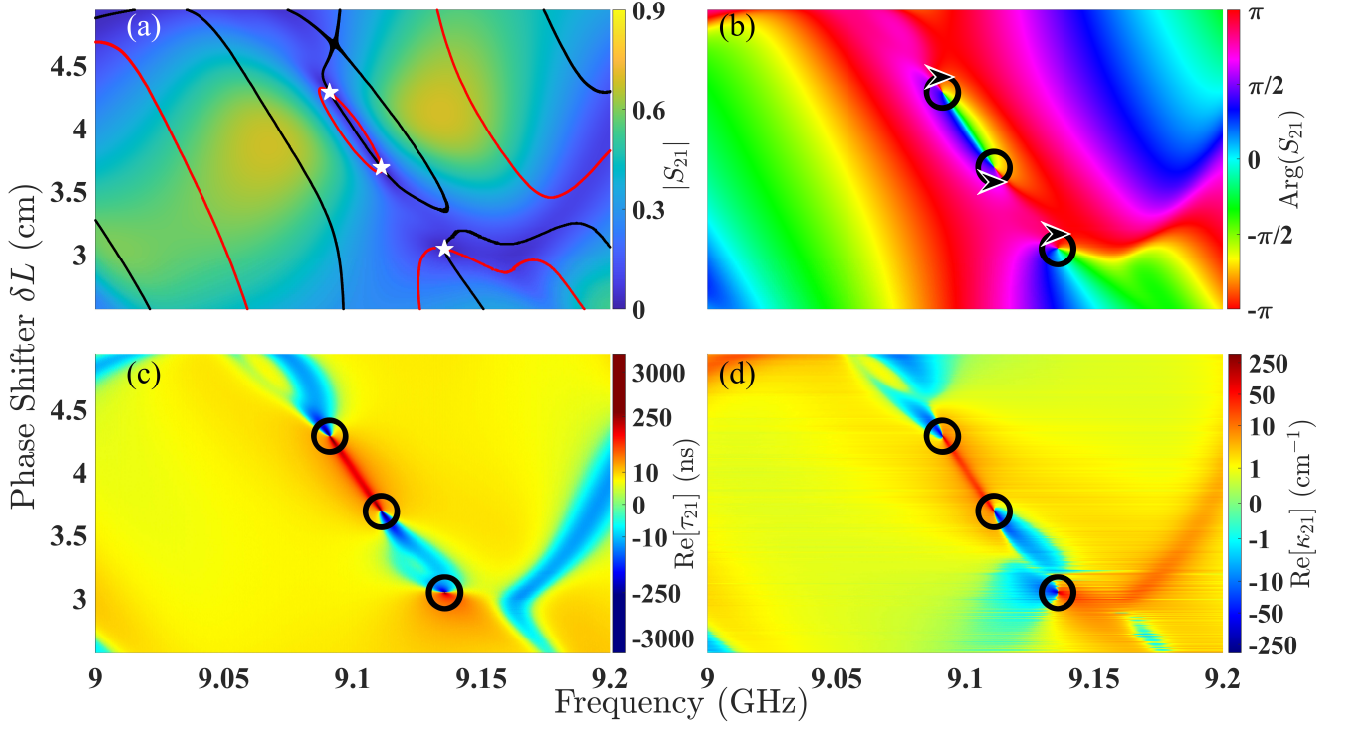}
\caption{Speckle pattern of Transmissionless Scattering Modes in $d=2$ parameter space. The experimental S-matrix data comes from a tetrahedral microwave graph. (a) Magnitude of $S_{21}$, red lines mark locations of $\text{Re}[S_{21}]=0$ while black lines mark locations of $\text{Im}[S_{21}]=0$. White symbols highlight intersections which correspond to $S_{21}=0+i0$ which are TSMs. (b) Phase of $S_{21}$, black circles show locations of phase windings which align exactly with the white symbols in (a). (c) Complex transmission time delay $\tau_{21}$ defined as the frequency derivative of $\text{log}S_{21}$ with units of ns, divergences occur exactly at the TSMs marked in panels (a-b). (d) Complex transmission generalized delay $\kappa_{21}$ defined as the phase shifter length derivative $\text{log}S_{21}$ with units of cm$^{-1}$, divergences occur exactly at the TSMs marked in panels (a-b).}
\label{App_CTD_fig}
\end{figure*}

Every singularity can be associated with the divergence of a complex time delay. Examples \rev{include} Wigner-Smith time delay diverging at CPAs and transmission time delay diverging at TSMs \cite{Lei2020,Erb2024,shaibe2025,Kang2021,Huang2022}\rev{, with more examples given in Table.~\ref{CTDTable}}. New time delays can be defined for singularities which lack one in the literature. As an example, we will introduce a complex coalescence time delay which diverges at EP-2s in $M=2$ port non-Hermitian scattering systems. \rev{Scattering matrix} Exceptional points occur when $\lambda_1-\lambda_2=0$, which can be rewritten in terms of matrix invariants $\sqrt{(\text{Tr}S)^2-4\text{det}S}=0$. To avoid the square root of a complex number which leads to phase discontinuities, we square both sides:
\begin{equation}
    0=(\lambda_1-\lambda_2)^2=(\text{Tr}S)^2-4\text{det}S. \nonumber
\end{equation}

We then define the coalescence time delay as
\begin{equation}
    \tau_C=-i\frac{\partial}{\partial E}\text{log}[(\text{Tr}S)^2-4\text{det}S]. \label{CCTD_eqn}
\end{equation}

\begin{table*}[htbp]
    \centering
    \renewcommand{\arraystretch}{1.3} % Optional: improves row spacing
    \begin{tabular}{|P{4cm}|P{4cm}|P{5cm}|P{4cm}|}
    \hline
    \rev{\textbf{Singularity Name}} & \rev{\textbf{Complex Scalar Function That Goes To $0+i0$}} & \rev{\textbf{Phenomena}} & \rev{\textbf{Diverging Complex Time Delay}} \\
    \hline
    \rev{Coherent Perfect Absorption (CPA)} & \rev{$\text{det}S$} & \rev{Absolute absorption of all injected power} & \rev{$\tau_{WS}= \frac{-i}{M}\frac{\partial}{\partial E}\text{log}[\text{det}S]$} \\
    \hline
    \rev{Exceptional Point (EP-2)} & \rev{$\lambda_{j}-\lambda_{k}$} & \rev{Collapse of eigenbasis, square root splitting of eigenvalues} & \rev{$\tau_C=-i\frac{\partial}{\partial E}\text{log}[(\text{Tr}S)^2-4\text{det}S]$}\\
    \hline
    \rev{Reflectionless Scattering Mode (RSM)} & \rev{$S_{xx}$} & \rev{Zero reflected power at channel $x$} & \rev{$\tau_{xx}=-i\frac{\partial}{\partial E}\text{log}[S_{xx}]$} \\
    \hline
    \rev{Transmissionless Scattering Mode (TSM)} & \rev{$S_{xy}$} & \rev{Zero transmitted power from channel $y$ to channel $x$} & \rev{$\tau_{xy}=-i\frac{\partial}{\partial E}\text{log}[S_{xy}]$} \\
    \hline
    \rev{Reciprocal Transmission in Non-Reciprocal Systems} & \rev{$\delta T_{xy} = S_{xy}-S_{yx}$} & \rev{Lorentz symmetry in non-reciprocal system} & \rev{$\tau_{\delta T}=-i\frac{\partial}{\partial E}\text{log}[S_{xy}-S_{yx}]$} \\
    \hline
    \rev{Symmetric Reflection in Generic Systems} & \rev{$\delta R_{xy}=S_{xx}-S_{yy}$} & \rev{Reflection symmetry between channels $x$ and $y$ in non-symmetric system} & \rev{$\tau_{\delta R}=-i\frac{\partial}{\partial E}\text{log}[S_{xx}-S_{yy}]$} \\
    \hline
    \end{tabular}
    \caption{\rev{Table of fundamental singularities of non-unitary scattering matrices. Provided are the name of the singularity, the complex scalar function that goes to zero, a physical consequence of the singularity, and the complex time delay that diverges at that singularity.}}
    \label{CTDTable}
\end{table*}

The typical way to define time delay uses a derivative in energy (or equivalently frequency), but the delay operator can be generalized further as a derivative in $p$ where $p$ is any parameter of the scattering matrix \cite{Ambichl2017,Ambichl2018}. For example, a phase shifter length $\delta L$ for a graph system like in Figs.~\ref{Fig1} and \ref{App_3Para}, or a Hamiltonian perturbation in simulations like in Figs.~\ref{Fig5} and \ref{App_EpsLoss}.

% \begin{figure}[bht]
% \includegraphics[width=0.48\textwidth]{FigApp_CTD.png}
% \caption{Speckle pattern of Transmissionless Scattering Modes in $d=2$ parameter space. The experimental S-matrix data comes from a tetrahedral microwave graph. (a) Magnitude of $S_{21}$, red lines mark locations of $\text{Re}[S_{21}]=0$ while black lines mark locations of $\text{Im}[S_{21}]=0$. White symbols highlight intersections which correspond to $S_{21}=0+i0$ which are TSMs. (b) Phase of $S_{21}$, black circles show locations of phase windings which align exactly with the white symbols in (a). (c) Complex transmission time delay $\tau_{21}$ defined as the frequency derivative of $\text{log}S_{21}$ with units of ns, divergences occur exactly at the TSMs marked in panels (a-b). (d) Complex transmission generalized delay $\kappa_{21}$ defined as the phase shifter length derivative $\text{log}S_{21}$ with units of cm$^{-1}$, divergences occur exactly at the TSMs marked in panels (a-b).}
% \label{App_CTD_fig}
% \end{figure}

We demonstrate this in Fig.~\ref{App_CTD_fig} using the transmission element $S_{21}$ from an experimental graph. Panels (a-b) show the speckle pattern in magnitude and phase of the TSM singularities. Panel (c) has the real component of the typical transmission time delay defined as $\tau_{21}=-i\frac{\partial}{\partial E}\text{log}S_{21}$ with units of nanoseconds. It can be seen that the divergences of the transmission time delay, where the surface instantly changes from deep red to deep blue marked by the black circles, occur precisely at the TSMs and nowhere else. Finally, in panel (d) we plot the real part of a transmission generalized delay calculated by the formula
\begin{equation}
    \kappa_{21}=-i\frac{\partial}{\partial\delta L}\text{log}S_{21} \label{EQN_Kappa}
\end{equation}
which has units of inverse centimeters. The divergences in panel (d) again occur exactly at the singularities. The data in panel (d) isn't as clean as the other panels and horizontal streaks can be seen. This is because the phase shifter is not as stable a perturbation as frequency and there is a longer time taken between measurements in the vertical direction as opposed to the horizontal. 

The same result would be achieved by taking the derivative of any other function of the scattering matrix, and with respect to any parameter, though in the latter case the units may be different. A logical question then arises, what is the physical interpretation of a complex general delay that does not have units of seconds? As of yet we do not have a new application or understanding of generalized delay beyond its use to find singularities.

\section{Fundamental Singularities Enumerated} \label{App_listsings}

Below we provide two tables of many fundamental scattering singularities, Table.~\ref{SingTable_EST} listing well studied ones with known applications, while Table.~\ref{SingTable_New} has singularities unexplored in the literature or introduced for the first time in this paper. In both tables, we are considering generic wave systems with either no or minimal inherent symmetries. Otherwise certain singularities, such as symmetric reflection, are not rare events but required phenomena. This is not a complete list since any possible complex scalar function formed out of the scattering matrix would have singularities.

\begin{table*}[htbp]
    \centering
    \renewcommand{\arraystretch}{1.3} % Optional: improves row spacing
    \begin{tabular}{|P{5cm}|P{5cm}|P{5cm}|P{2cm}|}
    \hline
    \textbf{Singularity Name} & \textbf{Complex Scalar Function That Goes To $0+i0$} & \textbf{Phenomena} & \textbf{References} \\
    \hline
    Coherent Perfect Absorption (CPA) & $\text{det}S$ & Absolute absorption of all injected power & \cite{Chong2010,Fyodorov2017,Lei2020,Imani2020,Frazier2020,Hougne2021,Erb2024,Faul2025,Gupta2012,Pichler2019,Baranov2017,Song2014,Meng2017,IglesiasMartinez2025,Li2022,Vetlugin2021,Vetlugin2022_Resolution,Vetlugin2022_AHOM,erb2025robustwavesplittersbased} \\
    \hline
    Exceptional Point (EP-2) & $\lambda_{j}-\lambda_{k}$ & Collapse of eigenbasis, square root splitting of eigenvalues & \cite{Berry2004,El-Ganainy18,Alu19,Özdemir19,Ashida20,Ding2022,PhysRevLett.106.213901,PhysRevLett.112.203901,Hodaei2017,Chen2017,PhysRevLett.86.787,PhysRevE.69.056216,PhysRevLett.106.150403,Gao2015,Shi2016,Doppler2016,Stehmann_2004,Suntharalingam2023,erb2025} \\
    \hline
    Orthogonality Point & $\langle R_{j}|R_{k}\rangle$ & Orthogonal eigenbasis for a non-unitary $S$-matrix & \cite{erb2025} \\
    \hline
    Reflectionless Scattering Mode (RSM) & $S_{xx}$ & Zero reflected power at channel $x$ & \cite{Jiang2024,Sol2023,Faul2025,Ferise2022,Davy2023,Sweeney2020,Stone2021} \\
    \hline
    Transmissionless Scattering Mode (TSM) & $S_{xy}$ & Zero transmitted power from channel $y$ to channel $x$ & \cite{Faul2025,Huang2022,Kang2021,Asano2016,Genack2024} \\
    \hline
    Reciprocal Transmission in Non-Reciprocal Systems & $\delta T_{xy}$ & Lorentz symmetry in non-reciprocal system & \cite{shaibe2025} \\
    \hline
    Symmetric Reflection in Generic Systems& $\delta R_{xy}$ & Reflection symmetry between channels $x$ and $y$ in non-symmetric system & \cite{Fyodorov2019,Osman2020,shaibe2025} \\
    \hline
    Complex Transmission Time Delay Zero & $\tau_{xy}\rev{=-i\frac{\partial}{\partial E}\text{log}[S_{xy}]}$ & Minimal distortion of transmitted pulse & \cite{Asano2016,giovannelli2025} \\
    \hline

    \end{tabular}
    \caption{Table of fundamental singularities of non-unitary scattering matrices. Provided are the name of the singularity, the complex scalar function that goes to zero, a physical consequence of the singularity, and relevant references.}
    \label{SingTable_EST}
\end{table*}

\begin{table*}[htbp]
    \centering
    \renewcommand{\arraystretch}{1.3} % Optional: improves row spacing
    \begin{tabular}{|P{6cm}|P{6cm}|P{5cm}|}
    \hline
    \textbf{Singularity Name} & \textbf{Complex Scalar Function That Goes To $0+i0$} & \textbf{Phenomena} \\
    \hline
        Other Principal Invariant Zeros of $M\times M$ $S$-matrix & $I_n = \sum_{1\le j < k < \cdots < n < M} \lambda_{j} \lambda_{k} \cdots \lambda_{n}$ & Uncertain\\
    \hline
        Complex Conjugate $S$-matrix Eigenvalues &  $\lambda_{j}-\lambda^*_{k}$ & Uncertain\\
    \hline
        Opposite $S$-matrix Eigenvalues &  $\lambda_{j}+\lambda_{k}$ & Uncertain\\
    \hline
        Negative Complex Conjugate $S$-matrix Eigenvalues &  $\lambda_{j}+\lambda^*_{k}$ & Uncertain\\
    \hline
        Equal Reflection and Transmission& $S_{xx}-S_{xy}$ & Reflection at channel $x$ equals transmission from channel $y$ to $x$\\
    \hline
    Complex Wigner-Smith Time Delay Zero & $\tau_{WS} \rev{= \frac{-i}{M}\frac{\partial}{\partial E}\text{log}[\text{det}S]}$ & Uncertain\\
    \hline
    Complex Reflection Time Delay Zero & $\tau_{xx}\rev{ = -i\frac{\partial}{\partial E}\text{log}[S_{xx}]}$ & Minimal distortion of reflected pulse\\
    \hline
    Complex Reflection Time Delay Difference Zero & $\tau_{\delta R}\rev{=-i\frac{\partial}{\partial E}\text{log}[S_{xx}-S_{yy}]}$ & Uncertain\\
    \hline
    Complex Transmission Time Delay Difference Zero & $\tau_{\delta T}\rev{=-i\frac{\partial}{\partial E}\text{log}[S_{xy}-S_{yx}]}$ & Uncertain\\
    \hline
        \textbf{Any} Complex Generalized Delay Zero (any parameter derivative of any function of the $S$-matrix) & $\kappa$ \rev{(See Eq.~\ref{EQN_Kappa} for an example)} & Uncertain\\
    \hline

    \end{tabular}
    \caption{Table of less explored fundamental singularities of non-unitary scattering matrices. Provided are the name of the singularity, the complex scalar function that goes to zero, and for some a physical consequence. These singularities all exist, but are not known to have applications yet.}
    \label{SingTable_New}
\end{table*}

%\clearpage
% \newpage 

%\renewcommand{\thefigure}{S\arabic{figure}}
\bigskip

%\textbf{MATERIALS AND METHODS} 

%\vspace{1cm}
%\appendix

%\section{\label{sec:Fig}Figures}  

% \newpage

% \textbf{EXTENDED DATA} 
%\vspace{1cm}

\clearpage

\bibliography{StatsTOP.bib}

%\bigskip
%\textbf{Author contributions}  S.M.A. conceived and directed the research, J.M.E. performed measurements, N.S. performed the simulations, measurements, and analysis.

%\bigskip
%\textbf{Competing interests}  The authors have no competing interests to declare. 

%\bigskip
%\textbf{Data and Materials Availability}  All data are available in the main text. 

%\bigskip
%\textbf{Correspondence}  and requests for materials should be addressed to Nadav Shaibe.

%\bigskip
% Recommended referees in no particular order:
% 1)	Douglas Stone, Yale University, douglas.stone@yale.edu
% 2)	Dmitry Savin, Brunel University of London, dmitry.savin@brunel.ac.uk
% 3)	Matthieu Davy, University of Rennes, matthieu.davy@univ-rennes.fr.
% 4)	Stefan Rotter, Vienna University of Technology, stefan.rotter@tuwien.ac.at
% 5)	Ulrich Kuhl, Côte d'Azur University, Ulrich.KUHL@univ-cotedazur.fr
% 6)	Azriel Genack, Queens College of CUNY & Chiral Photonics, azriel.genack@qc.cuny.edu
% 7)	Franco Nori, University of Michigan, nori@umich.edu

% We gratefully request the following people to be excluded from referee selection:
% 1)	Philipp del Hougne, University of Rennes, philipp.del-hougne@univ-rennes1.fr
% 2)	Barbara Dietz, Institute for Basic Science, South Korea, barbara@ibs.re.kr

\typeout{get arXiv to do 4 passes: Label(s) may have changed. Rerun}
\end{document}